# Measuring chemical evolution and gravitational dependence of $\alpha$ using ultraviolet Fe V and Ni V transitions in white-dwarf spectra

A. Ong, J. C. Berengut, and V. V. Flambaum

*School of Physics, University of New South Wales, Sydney, NSW 2052, Australia*

(Dated: October 10, 2013)

In this paper, we present the details of the *ab initio* high-precision configuration interaction and many-body perturbation theory calculations that were used in [1] to place limits on the dependence of the fine-structure constant, $\alpha$, on the gravitational field of the white-dwarf star G191-B2B. These calculations were combined with laboratory wavelengths and spectra from the Hubble Space Telescope Imaging Spectrograph to obtain the result $\Delta\alpha/\alpha = (4.2 \pm 1.6) \times 10^{-5}$ and $(-6.1 \pm 5.8) \times 10^{-5}$ using Fe V and Ni V transitions, respectively. The uncertainty in these results are dominated by the uncertainty in the laboratory wavelengths. In this work we also present *ab initio* calculations of the isotopic shifts of the Fe V transitions. We show that improved laboratory spectra will enable determination of the relative isotope abundances in Fe V to an accuracy $\sim 20\%$. Therefore this work provides a strong motivation for new laboratory measurements.

## I. INTRODUCTION

The light scalar fields that populate many modern theories of high-energy physics can change parameters of the standard model such as fundamental coupling constants [2]. Near massive bodies the effect of the scalar field can change since, like gravitational charge, the scalar charge is purely additive. Moreover, the strength of the coupling can increase or decrease near massive gravitating bodies depending on the theoretical model used [3]. For small variations in the fine-structure constant $\alpha = e^2/\hbar c \approx 1/137$, the dependence on the dimensionless gravitational potential $\phi = GM/rc^2$ can be approximately described by the linear relationship [2]

$$\frac{\delta\alpha}{\alpha} \equiv k_\alpha \Delta\phi = k_\alpha \Delta\left(\frac{GM}{rc^2}\right). \quad (1)$$

Here $k_\alpha$ is a sensitivity parameter [4]. This dependence has been investigated in certain theories of varying $\alpha$, where $\alpha$ is allowed to either increase ($\Delta\alpha/\alpha > 0$) or decrease ($\Delta\alpha/\alpha < 0$) on approach to a massive object [5–8], depending on the balance between electrostatic and magnetic energy in the ambient matter fields [3].

The strongest limit on $k_\alpha = (-5.5 \pm 5.2) \times 10^{-7}$ was obtained by comparing the radio-frequency transitions of two nearly-degenerate, opposite-parity excited states in atomic dysprosium [9]. This sensitivity is derived purely from annular changes in the gravitational potential ($\sim 3\%$) due to the ellipticity of the Earth's orbit around the Sun and the high precision of atomic clocks. The peak-to-trough sinusoidal change in the potential has magnitude $\Delta\phi = 3 \times 10^{-10}$.

A much stronger change in potential can be realised in the atmosphere of white-dwarf stars, where $\Delta\phi$ is 5 orders of magnitude larger than in laboratory-based experiments. This allows us to probe nonlinear coupling of $\Delta\alpha/\alpha$ on $\Delta\phi$. The distance between source and probe is $\sim 10^4$ times smaller than 1 AU, so we can also test the effect of, for example, a Yukawa-like scalar field $\Phi \sim e^{-mr}/r$ where $m$ is the mass of the light scalar.

High-resolution spectra of $Fe^{4+}$ and $Ni^{4+}$ 4s-4p transitions in the nearby ($\approx 45$ pc [10]), hot, hydrogen-rich (DA) white-dwarf star G191-B2B has been taken using the Hubble Space Telescope Imaging Spectrograph (STIS) [11]. G191-B2B has a mass of $M = 0.51 M_\odot$ and a radius of $R = 0.022 R_\odot$, thus the gravitational potential for ions in the atmosphere of this white dwarf relative to the laboratory is $\Delta\phi \approx 4.91 \times 10^{-5}$. The linear sensitivity parameter $k_\alpha = 0.7 \pm 0.3$ obtained from the analysis [1] is not competitive with the limits obtained from atomic clocks, but the white-dwarf result probes a very different field-strength regime.

It is instructive to compare the method with previous studies looking at cosmological variation of $\alpha$ using quasar absorption spectra [12–17]. In many ways, we can expect higher sensitivity from the white-dwarf spectra than the quasar absorption spectra [1] (although we stress again that the fundamental physics being tested in these two experiments is quite different). Firstly there are many more resolved lines per source (around 100 in $Fe^{4+}$ and 50 in $Ni^{4+}$) which provides a statistical advantage over the typical few narrow lines per absorber in quasar absorption systems. Secondly, the ionization degrees of the ions available are larger, and thus they are more sensitive to $\alpha$-variation. This is because the sensitivity of an external electron to $\alpha$-variation, $q$, can be approximated using [18]

$$q \approx -I_n \frac{(Z\alpha)^2}{\nu(j+1/2)}. \quad (2)$$

Here $I_n$ is the ionization energy of the electron (in atomic units; 1Ry = 1/2), $Z$ is the bare nuclear charge, $\nu$ is the non-integer effective principal quantum number and $j$ is the angular momentum of the external electron. Taken together, from these two advantages we would expect the value of $\Delta\alpha/\alpha$ extracted from the white-dwarf study to be more than an order-of-magnitude more sensitive (per system) than the quasar studies, reaching statistical accuracies below $10^{-6}$. However, the work is limited by relatively poor laboratory wavelengths.

In this paper, we present our calculations of $q$-values for the Fe V and Ni V transitions used in [1]. In addition, we also present an additional important motivation for remeasurement of their laboratory wavelengths. We calculate the isotope shifts in Fe V, which can be used to measure the relative isotope abundance of Fe around G191-B2B and similar white-dwarf stars. These isotope abundances can then be used to constrain theories of galactic chemical evolution (see, e.g., [19]). We show that with improved laboratory wavelengths, we could constrain the isotopic abundance ratios in Fe to around $\sim 20\%$. The method and results of our isotope shift calculations are presented in Section III.

## II. ENERGY CALCULATION AND RELATIVISTIC SHIFT

When expressed in atomic energy units, the non-relativistic transition energies are insensitive to $\alpha$-variation. However, the relativistic shifts of atomic energy levels are $\alpha$-dependent. We parametrize the sensitivity of a transition to potential $\alpha$-variation using the $q$-value, defined by

$$q = \frac{d\omega}{dx}\bigg|_{x=0}, \qquad (3)$$

where $x = (\alpha/\alpha_0)^2 - 1$ is the relative change of $\alpha^2$ from its present-day value of $\alpha_0^2$. Note that $\omega$ here is in atomic units: the non-relativistic dependence on $\alpha$ cancels when comparing two transitions. In our calculations, $q$ is obtained by taking the gradient of fully-relativistic energies with respect to $\alpha$ for a particular configuration by varying the value of $\alpha$ in the calculation. For our Fe$^{4+}$ calculation we used five values of $x = -0.002, -0.001, 0.0, 0.001,$ and $0.002$ to extract $q$, while in Ni$^{4+}$ we used $x = -0.01, 0.0,$ and $0.01$ to obtain a three point fit.

Below we outline our calculations, implementing the CI+MBPT method: configuration interaction with many-body perturbation theory corrections. Full details of the theory behind this method is presented in [20] (more detail on our particular implementation are provided in [21]). Briefly, our calculations begin by computing the Dirac-Fock (DF) energies of core orbitals in a chosen starting potential. This provides a "frozen-core" potential from which to generate a basis and perform CI. Our CI basis is formed from a set of between 40 to 70 $B$-splines [22, 23], which we diagonalise over the DF potential and select those orbitals with the lowest energy eigenvalue. A larger selection of these virtual orbitals are used to form many-body perturbation theory (MBPT) corrections to the potential, $\Sigma$.

The ions Fe$^{4+}$ and Ni$^{4+}$ are four and six valence electron ions respectively. The use of the CI+MBPT method to obtain energies, $q$-values, and isotope shifts in many-valence-electron systems is justified by past work (see, for example, Cr II [24]). Due to the large number of valence electrons, the size of the CI matrix grows beyond the limit of available computer memory before a fully converged result can be reached. This is particularly true for Ni$^{4+}$ which has two additional valence electrons over Fe$^{4+}$. Therefore, it becomes necessary to truncate the calculation and/or adjust parameters in order to make the calculation manageable. We used the available experimental data [25] as a benchmark to judge the "correctness" of calculations and thus decide on the most suitable set of parameters.

The DF calculation that we start with allows for a few different sensible choices of starting potentials (mean-field of all electrons). In a completely converged CI calculation, the effect of choosing the starting potential will be negligible – however, the choice of potential can make a significant difference for a truncated computation. As a specific example, in Fe$^{4+}$ we could reasonably expect to choose between either a $V^N$ or a $V^{N-1}$ starting potential by varying the number of electrons to be used. Furthermore, after choosing either one of these, we are free to choose the distribution of the electrons amongst available valence orbitals. For Fe$^{4+}$, we compared the results obtained using $3d^24s^2$ ($V^N$), $3d^4$ ($V^N$), and $3d^3$ ($V^{N-1}$) potentials and found those produced from a $3d^4$ $V^N$ potential to be the best. However, we note that there was good agreement between calculations performed using the three potentials. A similar result was obtained with Ni$^{4+}$ with the $3d^6$ $V^N$ potential found to give results closest to experimental data.

The next set of parameters we can vary affect the CI calculation. It consists of the leading configurations, the number of electron excitations taken from these leading configurations, and the number of virtual orbitals included at which the CI valence basis is truncated for the generation of the electron excitations. The number of leading configurations included significantly influences the overall size of the CI matrix. In Fe$^{4+}$, we tested the ($3d^4$, $3d^24s^2$) and the ($3d^4$, $3d^34s$, $3d^34p$) sets and found the latter to give better results. Again, the results produced by the other set of leading configurations were still reasonably close to the experimental results. In Ni$^{4+}$, the ($3d^6$, $3d^54s$, $3d^54p$) leading configuration set was found to give the best agreement with experiment, although we suspect that the ($3d^6$, $3d^54s$, $3d^54p$, $3d^44s^2$, $3d^44p^2$) superset of leading configurations would have improved agreement if not for computer memory limitations.

For the number of electron excitations, we can choose to either include a large number of single electron excitations from the leading configurations at the cost of some double electron excitations by choosing different CI valence basis sets for each; or include single and double electron excitations to the same CI valence basis. In Fe$^{4+}$, we were able to include both single and double electron excitations to the $11spdf7g$ valence orbitals, and this provided a better agreement with experiment than including single electron excitations to $15spdf$ and double electron excitation to $9spdf$. The resulting energy levels in both these cases are within 5% of their experimental values [25].

In Ni$^{4+}$, the number of valence orbitals we can use for the CI calculation is markedly smaller due to the two additional valence electrons. In this case, our calculations reveal that taking a large number of single electron excitations to 12$spdf$ and a smaller number of double excitations to 5$spdf$ produced a better result than to take single and double excitations to the 7$s$6$pdf$ level. Once again, the levels energies are within 5% of their experimental values.

Many-body perturbation theory (MBPT) corrections using a valence basis of 35$spdfghi$ for Fe$^{4+}$ and 30$spdfgh$ for Ni$^{4+}$ were then added to the CI calculation. We included one-body, two-body, and three-body MBPT diagrams to second-order in the residual Coulomb interaction. The final energies were improved by the addition of MBPT; agreement with experimental values were at the level $\sim 2\%$. Tables II and III illustrate the effect of the addition of MBPT. The $q$-values are stable with the addition of MBPT corrections – one can be certain that the MBPT $q$-values would differ by at most 5% (see Table II) from their CI-only values when differences in configuration mixing are not significant.

*Error estimates and state mixing:* In the results we present, some energy levels that were mixed in the CI calculation (that is, states that include a significant percentage of more than one term) become separated in the CI+MBPT calculation and vice versa. In such cases, the presented $g$-factors of the two calculations differ significantly. This can pose an issue when the $q$ values for the two levels are different – a strong mixing would imply the $q$ values of the two levels move towards the average of the two $q$ values, whereas less mixed states would have rather different $q$ values. In the absence of additional information, we are unable to decide which of these cases more closely represents the physical reality of the picture. We have marked with asterisks (*) in Table IV, levels that may have underestimated errors in $q$ as a result of the uncertainty in the true physical mixing.

For levels that do not demonstrate notable differences in mixing between calculations, we estimate our uncertainties in $q$ using the difference in values computed using CI and CI + MBPT for energy levels. The errors in $q$ values in Table IV were computed similarly by taking the difference between the final CI and CI + MBPT $q$ values of the transitions.

## III. ISOTOPE SHIFTS

The addition or removal of neutrons to the nucleus causes both a change in the nuclear mass and the nuclear charge distribution, both which in turn cause measurable shifts in the position of energy levels. These shifts can be parametrized, for two isotopes with with mass number $A$ and $A'$, using the equation

$$\delta\nu_{A,A'} = \left(\frac{1}{A} - \frac{1}{A'}\right)(k_{\mathrm{NMS}} + k_{\mathrm{SMS}}) + F\delta\langle r^2\rangle_{A,A'}, \quad (4)$$

where $k_{\mathrm{NMS}}$ is the normal mass shift constant, $k_{\mathrm{SMS}}$ is the specific mass shift constant and $F$ is the field shift constant. $k_{\mathrm{NMS}}$ is related to the transition frequency $\nu$ by

$$k_{\mathrm{NMS}} = -\frac{\nu}{1823}. \quad (5)$$

$k_{\mathrm{SMS}}$ is more difficult to obtain; it can be calculated through the addition of a new term to the Coulomb potential and varying a parameter related to the nuclear inverse mass, as described below. $F$ can be similarly obtained by instead varying the nuclear charge radius. Our calculations reveal that $k_{\mathrm{SMS}}$ can be quite different with the inclusion of many-body perturbation theory corrections, as seen in Table III. On the other hand, $F$ does not change much between CI and CI+MBPT calculations.

*Finite-field method:* To calculate $k_{\mathrm{SMS}}$, the all-order finite-field method is used [21, 24, 26–28]. A two-body SMS operator is added to the Coulomb potential to form an effective potential which can be written as

$$\tilde{Q} = \frac{1}{|\mathbf{r}_1 - \mathbf{r}_2|} + \lambda \mathbf{p}_1 \cdot \mathbf{p}_2 \quad (6)$$

and the CI+MBPT energy calculation is performed for several values of $\lambda$. The specific mass shift constant is then calculated using

$$k_{\mathrm{SMS}} = \left.\frac{d\omega}{d\lambda}\right|_{\lambda=0}. \quad (7)$$

The operator $\tilde{Q}$ has the same symmetry and structure as the ordinary Coulomb operator [26]. Our results were obtained by performing a five-point fit using values of $\lambda = -0.002, -0.001, 0.0, 0.001, 0.002$.

The field-shift constant $F$ can be calculated using one of two methods. The first approach is to vary the nuclear radius $R$ in the finite nuclear potential directly which in turn varies the mean squared charge radius, $\langle r^2 \rangle$. We perform the full CI+MBPT calculation for each value of $R$ and extract $F$ using

$$F = \frac{d\omega}{d\langle r^2\rangle}. \quad (8)$$

We used five values of $\langle r^2 \rangle$ in order to ensure linearity of the $F$ with respect to $\langle r^2 \rangle$.

The second approach used is similar to the finite-field method described for calculating $k_{\mathrm{SMS}}$. A finite perturbation $\delta U$ is added to the nuclear potential:

$$\delta U(r) = \lambda \left(U(R + \delta R, r) - U(R, r)\right), \quad (9)$$

where $R$ is the nuclear radius and $U(R, r)$ is the nuclear potential (e.g. a Fermi distribution with finite thickness parameter). The change in nuclear potential corresponds to a change in mean nuclear charge radius $\delta\langle r^2\rangle_U$, and we can extract the field-shift constant by calculating $\omega$ for several values of $\lambda$:

$$F = \frac{1}{\delta\langle r^2\rangle_U} \cdot \left.\frac{d\omega}{d\lambda}\right|_{\lambda=0}.$$

TABLE I. Isotopes of Fe.

| Isotope | Mass (amu) | Charge Radius (fm) | Abundance (%) |
|---------|------------|--------------------|---------------| 
| $^{54}$Fe | 53.9396 | 3.6931 (18) | 5.845 |
| $^{56}$Fe | 55.9349 | 3.7371 (15) | 91.754 |
| $^{57}$Fe | 56.9354 | 3.7534 (17) | 2.119 |
| $^{58}$Fe | 57.9333 | 3.7748 (14) | 0.282 |

Our calculations using the two methods were consistent within 1%.

*B-spline basis set:* The calculation of the field shift constant, $F$, is sensitive to the behaviour of the wavefunction at the origin. The B-spline basis of the Notre Dame (ND) group [23] was compared with B-splines generated using the dual kinetic-balance (DKB) finite basis set method [29, 30]. In our calculations, we found that the ND B-splines were unstable when used to calculate $F$, leading to inconsistencies when the size of the set of valence states used in the configuration interaction calculation was varied. The DKB B-splines, on the other hand, produced consistent values of $F$ regardless of the parameters used in the calculation. The ND calculation only agreed with the DKB calculation after the inclusion of a larger number of B-splines and a readjustment of the grid used.

This result reinforces the findings in [29], where the DKB B-Splines were found to have a behaviour at the origin that is closer to the physical reality. The original authors found that they required fewer B-splines when the DKB method was used compared to the ND method, a result our calculations also agree with.

*Error estimate:* The error in the total isotope shift, $\Delta\delta\nu_{A,A'}$, is largely given by the uncertainty in the specific mass shift constant. We estimate this uncertainty by using the difference between the CI and the CI + MBPT calculation as an upper bound, which can be obtained using Table III. The nuclear radii of the various isotopes and their errors are given in Table I.

## IV. DISCUSSION

Using our calculations for the relativistic shifts $q$, and isotopic shifts $\delta\nu_{A,A'}$ in Table III (for energy levels), we present a summary table containing these shifts for transitions observed in white-dwarf stars using the Hubble Space Telescope (HST) Space Telescope Imaging Spectrograph (STIS) in Table IV. For these observed lines, the extent of $\alpha$-variation is found to be $\Delta\alpha/\alpha = (4.2 \pm 1.6) \times 10^{-5}$ for Fe$^{4+}$. Furthermore, a limit on the deviation of isotopic abundances in white-dwarf stars from their terrestrial values can be placed using the differences between their observed and terrestrial wavelengths, in conjunction with the information in Table IV. In both the astronomical data and our reference laboratory wavelengths, the spectra of individual isotopes are not resolved. Thus, the available wavelength values are averages weighted by the isotopic abundances. For a particular white-dwarf line with wavelength $\lambda_{\text{obsv}}$ observed from Earth containing isotopes with mass numbers $A$ and $A'$, we can write

$$\lambda_{\text{obsv}} = (1 + z)\left[(1 - P_{A'})\lambda_A + P_{A'}\lambda_{A'}\right] , \quad (10)$$

where $z$ is the redshift, $P_{A'}$ is the isotopic abundance of isotope $A'$, and $\lambda_A$ and $\lambda_{A'}$ are the transition wavelengths assuming 100% abundance of isotopes $A$ and $A'$ respectively. Using Eq. (4), we may write this as

$$\frac{\Delta\lambda}{\lambda_A} = \frac{\lambda_{\text{obsv}} - \lambda_A}{\lambda_A} = z - P_{A'}\frac{\delta\omega_{A',A}}{\omega_A}(1 + z) . \quad (11)$$

Plotting $\Delta\lambda/\lambda_A$ against $\delta\omega_{A',A}/\omega_A \equiv Q_{\text{IS}}$ allows us to extract the abundance of isotope $A'$ immediately. The range of $Q_{\text{IS}}$ for Fe$^{4+}$ available from our calculations spans an interval of $\sim 6 \times 10^{-7}$. However, the errors in the laboratory wavelengths limits the accuracy of $\Delta\lambda/\lambda$ to $\sim 1 \times 10^{-5}$, thus rendering it insufficient to determine the isotopic abundances in this case. On the other hand, a laboratory accuracy of $\sim 1 \times 10^{-7}$ would allow for a determination of $P_{A'}$ to a level of $\sim 20\%$.

The observation of $3d$-$4p$ transitions may be another option for determining the isotope abundance. Such transitions would have energy intervals in the range of $(150000, 350000)$ cm$^{-1}$ corresponding to wavelengths in the range of $(30, 65)$ nm. These wavelength ranges are not within the range of observable wavelengths of the STIS. Nevertheless, we calculate the normal mass shifts to be in the range of $(-4000, -2500)$ GHz amu, while the specific mass shifts are in the range of $(10000, 11000)$ GHz amu. The corresponding volume shifts are on the order of $\sim 600$ MHz fm$^{-2}$, making them insignificant for these transitions. The total shift between the $^{56}$Fe and $^{58}$Fe is then about $\delta\nu_{58,56} \sim -5$ GHz. Compared to the $4s$-$4p$ transitions, this means that $\delta\nu_{3d-4p} \sim -5\delta\nu_{4s-4p}$, while $\nu_{3d-4p} \sim 2\nu_{4s-4p}$. Thus $\frac{\delta\nu}{\nu}\big|_{3d-4p} \sim -2.5\frac{\delta\nu}{\nu}\big|_{4s-4p}$. Therefore, these transitions would therefore allow for the accurate determination of isotopic abundances when combined with the $4s$-$4p$ transitions presented here.

## V. CONCLUSION

In closing, we have used spectral data from a white dwarf system to explore the possible coupling of the variation of fundamental constants to a strong gravitational potential. Furthermore, we showed how the differences in isotopic abundances relative to terrestrial abundances may be extracted. This information can be used to constrain models of the chemical evolution of our galaxy. Finally, we note that our results are limited by the uncertainties in the known terrestrial wavelengths of the transitions in both Fe$^{4+}$ and Ni$^{4+}$, with much tighter limits possible with accurate laboratory data. Our work therefore provides a strong motivation for new laboratory measurements.

TABLE II: Energy levels relative to the $3d^4\ ^5D_0$ ground state and Landé $g$-factors of $4s$ and $4p$ levels in $Fe^{4+}$. Quantities labelled CI were obtained using configuration interaction, while quantities labelled CI + $\Sigma$ were obtained using configuration interaction plus many-body perturbation theory.

| Level | | Energy (cm$^{-1}$) | | | $g$-factor | | |
|---|---|---|---|---|---|---|---|
| | | CI | CI +$\Sigma$ | Exp. [25] | CI | CI +$\Sigma$ | LS |
| $3d^4$ | $^5D_0$ | 0 | 0 | 0 | 0 | 0 | |
| $3d^4$ | $^5D_1$ | 159 | 165 | 142 | 1.4999 | 1.4999 | 1.5 |
| $3d^4$ | $^5D_2$ | 468 | 484 | 417 | 1.4998 | 1.4998 | 1.5 |
| $3d^4$ | $^5D_3$ | 912 | 942 | 803 | 1.4997 | 1.4996 | 1.5 |
| $3d^4$ | $^5D_4$ | 1475 | 1523 | 1283 | 1.4995 | 1.4995 | 1.5 |
| $3d^4$ | $^3P2_0$ | 28652 | 27027 | 24055 | | | |
| $3d^4$ | $^3H_4$ | 26985 | 26756 | 24933 | 0.8031 | 0.8061 | 0.8 |
| $3d^4$ | $^3P2_1$ | 29590 | 28043 | 24973 | 1.4988 | 1.4981 | 1.5 |
| $3d^4$ | $^3H_5$ | 27335 | 27129 | 25226 | 1.0348 | 1.0359 | 1.033 |
| $3d^4$ | $^3H_6$ | 27721 | 27541 | 25529 | 1.1662 | 1.1661 | 1.167 |
| $3d^4$ | $^3P2_2$ | 31229 | 29733 | 26468 | 1.4980 | 1.4978 | 1.5 |
| $3d^4$ | $^3F2_2$ | 31055 | 29397 | 26761 | 0.6684 | 0.6681 | 0.667 |
| $3d^4$ | $^3F2_3$ | 31139 | 29462 | 26842 | 1.0658 | 1.0617 | 1.083 |
| $3d^4$ | $^3F2_4$ | 31315 | 29625 | 26974 | 1.2373 | 1.2337 | 1.25 |
| $3d^4$ | $^3G_3$ | 34027 | 32147 | 29817 | 0.7681 | 0.7723 | 0.75 |
| $3d^4$ | $^3G_4$ | 34434 | 32564 | 30147 | 1.0588 | 1.059 | 1.05 |
| $3d^4$ | $^3G_5$ | 34753 | 32895 | 30430 | 1.1985 | 1.1974 | 1.2 |
| $3d^4$ | $^1G2_4$ | 43009 | 39814 | 36586 | 1.0014 | 1.0018 | 1 |
| $3d^4$ | $^3D_3$ | 43968 | 39668 | 36630 | 1.3327 | 1.3327 | 1.333 |

TABLE II – *Continued from previous page*

| Level | | Energy (cm$^{-1}$) | | | g-factor | | |
|---|---|---|---|---|---|---|---|
| | | CI | CI +$\Sigma$ | Exp. [25] | CI | CI +$\Sigma$ | LS |
| $3d^4$ | $^3D_2$ | 44079 | 39809 | 36759 | 1.1661 | 1.167 | 1.167 |
| $3d^4$ | $^3D_1$ | 44224 | 39937 | 36925 | 0.5015 | 0.5023 | 0.5 |
| $3d^4$ | $^1I_6$ | 40841 | 40241 | 37512 | 1.0005 | 1.0006 | 1 |
| $3d^4$ | $^1S2_0$ | 49881 | 43418 | 39633 | | | |
| $3d^4$ | $^1D2_2$ | 53243 | 51089 | 46291 | 1.0011 | 1.0009 | 1 |
| $3d^4$ | $^1F_3$ | 61424 | 57246 | 52733 | 1.0008 | 1.0008 | 1 |
| $3d^4$ | $^3P1_2$ | 72538 | 67756 | 61854 | 1.4991 | 1.499 | 1.5 |
| $3d^4$ | $^3F1_4$ | 72001 | 68224 | 62238 | 1.2494 | 1.2494 | 1.25 |
| $3d^4$ | $^3F1_2$ | 72032 | 68222 | 62321 | 0.6674 | 0.6674 | 0.667 |
| $3d^4$ | $^3F1_3$ | 72093 | 68307 | 62364 | 1.0830 | 1.0830 | 1.083 |
| $3d^4$ | $^3P1_1$ | 73614 | 68866 | 62914 | 1.4998 | 1.4997 | 1.5 |
| $3d^4$ | $^3P1_0$ | 74162 | 69431 | 63420 | | | |
| $3d^4$ | $^1G1_4$ | 81641 | 77831 | 71280 | 1.0006 | 1.0006 | 1 |
| $3d^4$ | $^1D1_2$ | 109739 | 102229 | 93832 | 1.0000 | 1.0000 | 1 |
| $3d^4$ | $^1S1_0$ | 141458 | 133273 | 121130 | | | |
| $3d^3(^4F)4s$ | $^5F_1$ | 194157 | 185985 | 186434 | 0.0011 | 0.0013 | 0 |
| $3d^3(^4F)4s$ | $^5F_2$ | 194503 | 186323 | 186726 | 0.9998 | 0.9999 | 1 |
| $3d^3(^4F)4s$ | $^5F_3$ | 195021 | 186831 | 187158 | 1.2496 | 1.2497 | 1.25 |
| $3d^3(^4F)4s$ | $^5F_4$ | 195703 | 187501 | 187719 | 1.3496 | 1.3496 | 1.35 |
| $3d^3(^4F)4s$ | $^5F_5$ | 196538 | 188318 | 188395 | 1.3995 | 1.3995 | 1.4 |
| $3d^3(^4F)4s$ | $^3F_2$ | 203002 | 195431 | 195196 | 0.6679 | 0.6680 | 0.667 |
| $3d^3(^4F)4s$ | $^3F_3$ | 203886 | 196280 | 195933 | 1.0835 | 1.0834 | 1.083 |
| $3d^3(^4F)4s$ | $^3F_4$ | 204993 | 197365 | 196839 | 1.2494 | 1.2492 | 1.25 |
| $3d^3(^4P)4s$ | $^5P_1$ | 216371 | 205239 | 204730 | 2.4909 | 2.4927 | 2.5 |
| $3d^3(^4P)4s$ | $^5P_2$ | 216641 | 205515 | 204975 | 1.8271 | 1.8283 | 1.833 |
| $3d^3(^4P)4s$ | $^5P_3$ | 217310 | 206157 | 205536 | 1.6660 | 1.6661 | 1.667 |
| $3d^3(^2G)4s$ | $^3G_3$ | 219245 | 209658 | 208838 | 0.7506 | 0.7507 | 0.75 |
| $3d^3(^2G)4s$ | $^3G_4$ | 219555 | 209965 | 209110 | 1.0478 | 1.0484 | 1.05 |
| $3d^3(^2G)4s$ | $^3G_5$ | 220043 | 210454 | 209524 | 1.1969 | 1.1976 | 1.2 |
| $3d^3(^4P)4s$ | $^3P_0$ | 224416 | 213930 | 212542 | 0 | 0 | |
| $3d^3(^4P)4s$ | $^3P_1$ | 224720 | 214180 | 212818 | 1.4624 | 1.4598 | 1.5 |
| $3d^3(^2G)4s$ | $^1G_4$ | 223887 | 214643 | 213534 | 0.9806 | 0.9855 | 1 |
| $3d^3(^4P)4s$ | $^3P_2$ | 225598 | 215038 | 213649 | 1.4981 | 1.4981 | 1.5 |
| $3d^3(^2P)4s$ | $^3P_2$ | 226717 | 216232 | 214526 | 1.4433 | 1.3843 | 1.5 |
| $3d^3(^2P)4s$ | $^3P_1$ | 226612 | 216267 | 214611 | 1.4364 | 1.2592 | 1.5 |
| $3d^34s$ | 0 | 226814 | 216626 | a | | | |
| $3d^3(^2D2)4s$ | $^3D_1$ | 228639 | 217569 | 215783 | 0.6570 | 0.8015 | 0.5 |
| $3d^3(^2D2)4s$ | $^3D_3$ | 229870 | 218368 | 216538 | 1.3336 | 1.3335 | 1.333 |
| $3d^3(^2D2)4s$ | $^3D_2$ | 229721 | 218549 | 216593 | 1.2219 | 1.2774 | 1.167 |
| $3d^3(^2H)4s$ | $^3H_4$ | 226810 | 217829 | 216779 | 0.8227 | 0.8175 | 0.8 |
| $3d^3(^2H)4s$ | $^3H_5$ | 226872 | 217917 | 216860 | 1.0359 | 1.0354 | 1.033 |
| $3d^3(^2H)4s$ | $^3H_6$ | 227205 | 218266 | 217123 | 1.1667 | 1.1667 | 1.167 |
| $3d^3(^2P)4s$ | $^1P_1$ | 231796 | 221770 | 219487 | 0.9522 | 0.9854 | 1 |
| $3d^3(^2D2)4s$ | $^1D_2$ | 233909 | 222735 | 220621 | 1.0081 | 1.0102 | 1 |
| $3d^3(^2H)4s$ | $^1H_5$ | 231287 | 222634 | 221305 | 1.0010 | 1.0008 | 1 |
| $3d^3(^2F)4s$ | $^3F_4$ | 248021 | 235784 | 233634 | 1.2498 | 1.2498 | 1.25 |
| $3d^3(^2F)4s$ | $^3F_3$ | 248221 | 235976 | 233849 | 1.0833 | 1.0833 | 1.083 |
| $3d^3(^2F)4s$ | $^3F_2$ | 248388 | 236145 | 234027 | 0.6674 | 0.6674 | 0.667 |
| $3d^3(^2F)4s$ | $^1F_3$ | 252318 | 240350 | 237730 | 1.0004 | 1.0004 | 1 |
| $3d^3(^4F)4p$ | $^5G_2^o$ | 262596 | 255796 | 254803 | 0.3375 | 0.3380 | 0.333 |
| $3d^3(^4F)4p$ | $^5G_3^o$ | 263260 | 256459 | 255399 | 0.9183 | 0.9187 | 0.917 |
| $3d^3(^4F)4p$ | $^5G_4^o$ | 264129 | 257322 | 256178 | 1.1507 | 1.1510 | 1.15 |
| $3d^3(^4F)4p$ | $^5G_5^o$ | 265195 | 258401 | 257138 | 1.2668 | 1.2670 | 1.267 |
| $3d^3(^4F)4p$ | $^5D_1^o$ | 265667 | 258647 | 257742 | 0.6126 | 0.5627 | 0.5 |
| $3d^3(^4F)4p$ | $^5D_2^o$ | 266043 | 259035 | 258129 | 1.3169 | 1.2966 | 1.5 |
| $3d^3(^4F)4p$ | $^5G_6^o$ | 266462 | 259704 | 258297 | 1.3329 | 1.3329 | 1.333 |
| $3d^3(^2D1)4s$ | $^3D_3$ | 276666 | 263247 | 258434 | 1.3333 | 1.3333 | 1.333 |
| $3d^3(^4F)4p$ | $^5D_0^o$ | 266262 | 259350 | 258620 | | | |
| $3d^3(^4F)4p$ | $^3D_2^o$ | 267212 | 260268 | 258629 | 1.1958 | 1.2216 | 1.167 |

TABLE II – *Continued from previous page*

| Level | | Energy (cm$^{-1}$) | | | g-factor | | |
|---|---|---|---|---|---|---|---|
| | | CI | CI +$\Sigma$ | Exp. [25] | CI | CI +$\Sigma$ | LS |
| $3d^3(^2D1)4s$ | $^3D_2$ | 276886 | 263430 | 258629 | 1.1661 | 1.1662 | 1.167 |
| $3d^3(^4F)4p$ | $^5D_3^o$ | 266638 | 259642 | 258680 | 1.4526 | 1.4456 | 1.5 |
| $3d^3(^2D1)4s$ | $^3D_1$ | 277045 | 263542 | 258770 | 0.5001 | 0.5001 | 0.5 |
| $3d^3(^4F)4p$ | $^5D_1^o$ | 266622 | 259702 | 258892 | 1.0781 | 1.1389 | 1.5 |
| $3d^3(^4F)4p$ | $^5D_4^o$ | 267420 | 260407 | 259345 | 1.4909 | 1.4889 | 1.5 |
| $3d^3(^4F)4p$ | $^5F_2^o$ | 268035 | 261065 | 259376 | 1.1482 | 1.1431 | 1 |
| $3d^3(^4F)4p$ | $^5F_3^o$ | 269624 | 262552 | 259955 | 1.3364 | 1.3368 | 1.25 |
| $3d^3(^4F)4p$ | $^5F_1^o$ | 268035 | 261065 | 259995 | 0.3115 | 0.3010 | 0 |
| $3d^3(^4F)4p$ | $^5F_4^o$ | 268609 | 261574 | 260521 | 1.3473 | 1.3504 | 1.35 |
| $3d^3(^4F)4p$ | $^5F_5^o$ | 269274 | 262279 | 261052 | 1.3863 | 1.3878 | 1.4 |
| $3d^3(^4F)4p$ | $^3D_3^o$ | 267884 | 260923 | 261180 | 1.2838 | 1.292 | 1.333 |
| $3d^3(^2D1)4s$ | $^1D_2$ | 280958 | 267772 | 262509 | 1.0003 | 1.0001 | 1 |
| $3d^3(^4F)4p$ | $^3G_3^o$ | 271508 | 265130 | 263899 | 0.7581 | 0.7567 | 0.75 |
| $3d^3(^4F)4p$ | $^3G_4^o$ | 272191 | 265776 | 264434 | 1.0600 | 1.0585 | 1.05 |
| $3d^3(^4F)4p$ | $^3G_5^o$ | 273082 | 266621 | 265113 | 1.2121 | 1.2103 | 1.2 |
| $3d^3(^4F)4p$ | $^3F_2^o$ | 274327 | 267991 | 266613 | 0.6699 | 0.6694 | 0.667 |
| $3d^3(^4F)4p$ | $^3F_3^o$ | 275101 | 268724 | 267240 | 1.0836 | 1.0834 | 1.083 |
| $3d^3(^4F)4p$ | $^3F_4^o$ | 275962 | 269561 | 267929 | 1.2491 | 1.2490 | 1.25 |
| $3d^3(^4P)4p$ | $^5P_1^o$ | 284981 | 275215 | 273643 | 2.4792 | 2.4772 | 2.5 |
| $3d^3(^4P)4p$ | $^5P_2^o$ | 285545 | 275753 | 274136 | 1.8216 | 1.8182 | 1.833 |
| $3d^3(^4P)4p$ | $^5D_0^o$ | 286600 | 276674 | 274753 | | | |
| $3d^3(^4P)4p$ | $^5P_3^o$ | 286333 | 276590 | 274930 | 1.6632 | 1.6631 | 1.667 |
| $3d^3(^4P)4p$ | $^5D_1^o$ | 286960 | 277063 | 275147 | 1.4913 | 1.4945 | 1.5 |
| $3d^3(^4P)4p$ | $^3P_2^o$ | 287234 | 277349 | 275374 | 1.4966 | 1.5017 | 1.5 |
| $3d^3(^4P)4p$ | $^3H_4^o$ | 286744 | 278504 | 276430 | 0.8136 | 0.8131 | 0.8 |
| $3d^3(^4P)4p$ | $^3P_0^o$ | 288302 | 278440 | 276435 | | | |
| $3d^3(^4P)4p$ | $^5D_2^o$ | 288604 | 278741 | 276759 | 1.5052 | 1.5081 | 1.5 |
| $3d^3(^4P)4p$ | $^3P_1^o$ | 288640 | 278794 | 276766 | 1.5087 | 1.5115 | 1.5 |
| $3d^3(^4P)4p$ | $^5D_3^o$ | 288842 | 279031 | 277069 | 1.4370 | 1.4982 | 1.5 |
| $3d^3(^2G)4p$ | $^3H_5^o$ | 287631 | 279446 | 277293 | 1.0456 | 1.0447 | 1.033 |
| $3d^3(^4P)4p$ | $^5D_4^o$ | 289939 | 280123 | 278076 | 1.4988 | 1.4993 | 1.5 |
| $3d^3(^2G)4p$ | $^3H_6^o$ | 289105 | 280926 | 278651 | 1.1651 | 1.1654 | 1.167 |
| $3d^3(^2G)4p$ | $^3G_3^o$ | 288901 | 280744 | 278794 | 0.8412 | 0.7940 | 0.75 |
| $3d^3(^2G)4p$ | $^3G_4^o$ | 289693 | 281543 | 279503 | 1.0457 | 1.0579 | 1.05 |
| $3d^3(^2G)4p$ | $^3G_5^o$ | 290205 | 282123 | 280040 | 1.1731 | 1.1773 | 1.2 |
| $3d^3(^2G)4p$ | $^3F_4^o$ | 290783 | 282411 | 280367 | 1.1014 | 1.1056 | 1.25 |
| $3d^3(^2P)4p$ | $^1S_0^o$ | 292453 | 282626 | a | | | |
| $3d^3(^2G)4p$ | $^3F_2^o$ | 291494 | 282867 | 280540 | 0.6773 | 0.6896 | 0.667 |
| $3d^3(^2G)4p$ | $^3F_3^o$ | 291442 | 283023 | 280832 | 1.0486 | 1.0369 | 1.083 |
| $3d^3(^2P)4p$ | $^3P_1^o$ | 293803 | 284774 | 281945 | 1.4790 | 1.4507 | 1.5 |
| $3d^3(^2G)4p$ | $^1G_4^o$ | 292432 | 284178 | 282038 | 1.0956 | 1.0957 | 1 |
| $3d^3(^2P)4p$ | $^3P_0^o$ | 294016 | 285106 | 282235 | | | |
| $3d^3(^4P)4p$ | $^5S_2^o$ | 294802 | 284659 | 282424 | 1.6071 | 1.9561 | 2 |
| $3d^3(^2G)4p$ | $^1F_3^o$ | 293527 | 284950 | 282572 | 1.0041 | 1.0055 | 1 |
| $3d^3(^2P)4p$ | $^1D_2^o$ | 294577 | 285404 | 282605[b] | 1.3162 | 1.0385 | 1 |
| $3d^3(^2G)4p$ | $^1H_5^o$ | 292926 | 285112 | 282872 | 1.0027 | 1.0020 | 1 |
| $3d^3(^2P)4p$ | $^3P_2^o$ | 295742 | 286616 | 283686 | 1.5446 | 1.4403 | 1.5 |
| $3d^3(^2P)4p$ | $^3D_1^o$ | 295327 | 286584 | 283754 | 0.5275 | 0.5457 | 0.5 |
| $3d^3(^2H)4p$ | $^3H_4^o$ | 294535 | 286730 | 284690 | 0.8471 | 0.8324 | 0.8 |
| $3d^3(^2H)4p$ | $^3H_5^o$ | 294522 | 286832 | 284791 | 1.0347 | 1.0340 | 1.033 |
| $3d^3(^2P)4p$ | $^3D_2^o$ | 296908 | 287861 | 284911 | 1.1360 | 1.1232 | 1.167 |
| $3d^3(^2H)4p$ | $^3H_6^o$ | 294976 | 287304 | 285196 | 1.1634 | 1.1633 | 1.167 |
| $3d^3(^2P)4p$ | $^3D_3^o$ | 297195 | 288233 | 285474 | 1.3051 | 1.2826 | 1.333 |
| $3d^3(^2D2)4p$ | $^1P_1^o$ | 297957 | 288775 | 285962 | 0.9834 | 1.0092 | 1 |
| $3d^3(^2D2)4p$ | $^3F_2^o$ | 298947 | 288727 | 286155 | 0.8283 | 0.9229 | 0.667 |
| $3d^3(^2P)4p$ | $^3S_1^o$ | 298298 | 289076 | 286188 | 1.6914 | 1.6573 | 2 |
| $3d^3(^4P)4p$ | $^3D_3^o$ | 298134 | 288944 | 286431 | 1.3082 | 1.2502 | 1.333 |
| $3d^3(^4P)4p$ | $^3D_1^o$ | 298847 | 289484 | 286855 | 0.8332 | 0.8470 | 0.5 |
| $3d^3(^4P)4p$ | $^3D_2^o$ | 298118 | 289527 | 286863 | 1.0689 | 1.0060 | 1.167 |

TABLE II – *Continued from previous page*

| Level | | Energy (cm$^{-1}$) | | | g-factor | | |
|---|---|---|---|---|---|---|---|
| | | CI | CI +$\Sigma$ | Exp. [25] | CI | CI +$\Sigma$ | LS |
| $3d^3(^2D2)4p$ | $^3F_3^o$ | 299436 | 289941 | 287110 | 1.1315 | 1.2089 | 1.083 |
| $3d^3(^2H)4p$ | $^3I_5^o$ | 297227 | 289714 | 287441 | 0.8476 | 0.8463 | 0.833 |
| $3d^3(^2D2)4p$ | $^3F_4^o$ | 300288 | 290528 | 287620 | 1.1417 | 1.2475 | 1.25 |
| $3d^3(^2H)4p$ | $^3I_6^o$ | 297970 | 290488 | 288167 | 1.0258 | 1.0259 | 1.024 |
| $3d^3(^2D2)4p$ | $^3D_1^o$ | 301168 | 291403 | 288670 | 0.5479 | 0.5373 | 0.5 |
| $3d^3(^2H)4p$ | $^3I_7^o$ | 299024 | 291599 | 289172 | 1.1429 | 1.1429 | 1.143 |
| $3d^3(^2D2)4p$ | $^3D_2^o$ | 302090 | 292210 | 289390 | 1.1780 | 1.1699 | 1.167 |
| $3d^3(^2H)4p$ | $^1G_4^o$ | 300530 | 291971 | 289546 | 1.1124 | 1.0034 | 1 |
| $3d^3(^2D2)4p$ | $^3D_3^o$ | 302829 | 292881 | 289913 | 1.2305 | 1.2901 | 1.333 |
| $3d^3(^2H)4p$ | $^1I_5^o$ | 299333 | 292356 | 290099 | 1.0065 | 1.0153 | 1 |
| $3d^3(^2P)4p$ | $^3P_2^o$ | 303091 | 293332 | 290408 | 1.4759 | 1.4836 | 1.5 |
| $3d^3(^2D2)4p$ | $^3P_1^o$ | 303347 | 293504 | 290584 | 1.4947 | 1.4846 | 1.5 |
| $3d^3(^2D2)4p$ | $^3P_0^o$ | 303732 | 293893 | 290903 | | | |
| $3d^3(^2D2)4p$ | $^1F_3^o$ | 303917 | 294076 | 291231 | 1.0519 | 0.9554 | 1 |
| $3d^3(^2H)4p$ | $^3G_5^o$ | 302388 | 294610 | 292288 | 1.1912 | 1.1822 | 1.2 |
| $3d^3(^2H)4p$ | $^1I_6^o$ | 301895 | 294661 | 292366 | 1.0033 | 1.0030 | 1 |
| $3d^3(^2H)4p$ | $^3G_4^o$ | 302649 | 294722 | 292431 | 1.0458 | 1.0475 | 1.05 |
| $3d^3(^2H)4p$ | $^3G_3^o$ | 302312 | 294871 | 292513 | 0.8122 | 0.8484 | 0.75 |
| $3d^3(^4P)4p$ | $^3S_1^o$ | 306062 | 297477 | 294644 | 1.9057 | 1.915 | 2 |
| $3d^3(^2D2)4p$ | $^1D_2^o$ | 307979 | 298897 | 295716 | 1.0088 | 1.0059 | 1 |
| $3d^3(^2P)4p$ | $^1P_1^o$ | 308019 | 299292 | 295973 | 1.0547 | 1.0665 | 1 |
| $3d^3(^2F)4p$ | $^3F_2^o$ | 316699 | 305486 | 302293 | 0.6779 | 0.6767 | 0.667 |
| $3d^3(^2F)4p$ | $^3F_3^o$ | 316778 | 305579 | 302377 | 1.0755 | 1.0779 | 1.083 |
| $3d^3(^2F)4p$ | $^3F_4^o$ | 317000 | 305844 | 302603 | 1.2392 | 1.2417 | 1.25 |
| $3d^3(^2F)4p$ | $^3G_3^o$ | 320036 | 309427 | 306194 | 0.7711 | 0.7686 | 0.75 |
| $3d^3(^2F)4p$ | $^3G_4^o$ | 320456 | 309892 | 306623 | 1.0589 | 1.0569 | 1.05 |
| $3d^3(^2F)4p$ | $^3G_5^o$ | 320929 | 310372 | 307064 | 1.1999 | 1.1999 | 1.2 |
| $3d^3(^2F)4p$ | $^3D_3^o$ | 321434 | 310780 | 307289 | 1.3155 | 1.3166 | 1.333 |
| $3d^3(^2F)4p$ | $^1D_2^o$ | 322198 | 311524 | 307644 | 1.0634 | 1.0641 | 1 |
| $3d^3(^2F)4p$ | $^3D_2^o$ | 322387 | 311756 | 308165 | 1.0933 | 1.0939 | 1.167 |
| $3d^3(^2F)4p$ | $^3D_1^o$ | 322827 | 312170 | 308672 | 0.5012 | 0.5011 | 0.5 |
| $3d^3(^2F)4p$ | $^1G_4^o$ | 324406 | 314671 | 311181 | 1.0020 | 1.0014 | 1 |
| $3d^3(^2F)4p$ | $^1F_3^o$ | 325128 | 315098 | 311539 | 1.0054 | 1.0042 | 1 |
| $3d^3(^2D1)4p$ | $^3D_1^o$ | 345129 | 333204 | 327534 | 0.5081 | 0.5092 | 0.5 |
| $3d^3(^2D1)4p$ | $^3D_2^o$ | 345197 | 333282 | 327605 | 1.1655 | 1.1652 | 1.167 |
| $3d^3(^2D1)4p$ | $^3D_3^o$ | 345512 | 333663 | 327924 | 1.3255 | 1.3238 | 1.333 |
| $3d^3(^2D1)4p$ | $^1D_2^o$ | 347377 | 335425 | 329849 | 0.9668 | 0.9472 | 1 |
| $3d^3(^2D1)4p$ | $^3F_2^o$ | 349286 | 337092 | 331334 | 0.7072 | 0.7285 | 0.667 |
| $3d^3(^2D1)4p$ | $^3F_3^o$ | 349341 | 337177 | 331367 | 1.0863 | 1.0887 | 1.083 |
| $3d^3(^2D1)4p$ | $^3F_4^o$ | 349982 | 337859 | 332017 | 1.2498 | 1.2497 | 1.25 |
| $3d^3(^2D1)4p$ | $^3P_2^o$ | 352735 | 340334 | 334509 | 1.4933 | 1.4918 | 1.5 |
| $3d^3(^2D1)4p$ | $^3P_1^o$ | 353518 | 341107 | 335268 | 1.4918 | 1.4907 | 1.5 |
| $3d^3(^2D1)4p$ | $^3P_0^o$ | 353898 | 341462 | 335643 | | | |
| $3d^3(^2D1)4p$ | $^1F_3^o$ | 353558 | 341962 | 335947 | 1.0043 | 1.0035 | 1 |
| $3d^3(^2D1)4p$ | $^1P_1^o$ | 360184 | 348790 | 342462 | 1.0000 | 1.0000 | 1 |

[a] This level has not been observed in the experimental data.
[b] This level was identified in the experimental data as $3d^3(^4P)4p\ ^5S_2^o$.

TABLE III: Calculated sensitivities to $\alpha$-variation $q$, specific mass shift constant $k_{\text{SMS}}$, and field shift constants $F$ (all relative to the ground state) of $4s$ and $4p$ energy levels in Fe$^{4+}$. Quantities labelled CI were obtained using configuration interaction, while quantities labelled CI + $\Sigma$ were obtained using configuration interaction plus many-body perturbation theory.

| Level | | $q$ (cm$^{-1}$) | | $k_{\text{SMS}}$ (GHz·amu) | | $F$ (MHz·fm$^{-2}$) | |
|---|---|---|---|---|---|---|---|
| | | CI | CI +$\Sigma$ | CI | CI +$\Sigma$ | CI | CI +$\Sigma$ |
| $3d^4$ | $^5D_0$ | 0 | 0 | 0 | 0 | -4298 | -3685 |
| $3d^4$ | $^5D_1$ | 172 | 179 | 8 | 7 | -4297 | -3684 |

TABLE III – *Continued from previous page*

| Level | | $q$ (cm$^{-1}$) | | $k_{\text{SMS}}$ (GHz·amu) | | $F$ (MHz·fm$^{-2}$) | |
|---|---|---|---|---|---|---|---|
| | | CI | CI +$\Sigma$ | CI | CI +$\Sigma$ | CI | CI +$\Sigma$ |
| $3d^4$ | $^5D_2$ | 496 | 517 | 22 | 20 | -4296 | -3683 |
| $3d^4$ | $^5D_3$ | 941 | 978 | 42 | 39 | -4294 | -3681 |
| $3d^4$ | $^5D_4$ | 1479 | 1533 | 66 | 65 | -4292 | -3679 |
| $3d^4$ | $^3P2_0$ | -940 | -977 | 335 | -131 | -4272 | -3669 |
| $3d^4$ | $^3H_4$ | 283 | 180 | 346 | 1451 | -4268 | -3645 |
| $3d^4$ | $^3P2_1$ | 67 | 129 | 380 | -80 | -4268 | -3665 |
| $3d^4$ | $^3H_5$ | 681 | 623 | 363 | 1474 | -4267 | -3643 |
| $3d^4$ | $^3H_6$ | 1082 | 1067 | 380 | 1505 | -4265 | -3641 |
| $3d^4$ | $^3P2_2$ | 1628 | 1734 | 452 | -8 | -4262 | -3659 |
| $3d^4$ | $^3F2_2$ | 516 | 586 | 413 | 499 | -4264 | -3659 |
| $3d^4$ | $^3F2_3$ | 490 | 518 | 413 | 530 | -4264 | -3659 |
| $3d^4$ | $^3F2_4$ | 638 | 679 | 419 | 545 | -4264 | -3659 |
| $3d^4$ | $^3G_3$ | 675 | 751 | 459 | 958 | -4260 | -3654 |
| $3d^4$ | $^3G_4$ | 1077 | 1146 | 477 | 982 | -4258 | -3652 |
| $3d^4$ | $^3G_5$ | 1307 | 1410 | 488 | 1022 | -4257 | -3651 |
| $3d^4$ | $^1G2_4$ | 1097 | 1276 | 582 | 971 | -4248 | -3656 |
| $3d^4$ | $^3D_3$ | 662 | 788 | 597 | 288 | -4250 | -3669 |
| $3d^4$ | $^3D_2$ | 754 | 929 | 601 | 293 | -4249 | -3668 |
| $3d^4$ | $^3D_1$ | 953 | 1102 | 610 | 300 | -4249 | -3668 |
| $3d^4$ | $^1I_6$ | 875 | 853 | 535 | 2211 | -4252 | -3625 |
| $3d^4$ | $^1S2_0$ | 948 | 1255 | 678 | -140 | -4240 | -3688 |
| $3d^4$ | $^1D2_2$ | 913 | 982 | 712 | 384 | -4244 | -3636 |
| $3d^4$ | $^1F_3$ | 669 | 781 | 819 | 747 | -4233 | -3643 |
| $3d^4$ | $^3P1_2$ | 187 | 324 | 973 | 174 | -4219 | -3629 |
| $3d^4$ | $^3F1_4$ | 652 | 774 | 984 | 698 | -4215 | -3612 |
| $3d^4$ | $^3F1_2$ | 711 | 799 | 989 | 701 | -4215 | -3611 |
| $3d^4$ | $^3F1_3$ | 828 | 933 | 993 | 704 | -4215 | -3611 |
| $3d^4$ | $^3P1_1$ | 1228 | 1395 | 1021 | 219 | -4215 | -3624 |
| $3d^4$ | $^3P1_0$ | 1767 | 1948 | 1046 | 244 | -4213 | -3623 |
| $3d^4$ | $^1G1_4$ | 691 | 779 | 1097 | 1208 | -4208 | -3601 |
| $3d^4$ | $^1D1_2$ | 531 | 721 | 1492 | 628 | -4176 | -3598 |
| $3d^4$ | $^1S1_0$ | 382 | 611 | 1932 | 302 | -4135 | -3548 |
| $3d^3(^4F)4s$ | $^5F_1$ | -6896 | -6690 | 11109 | 11489 | -1666 | -1024 |
| $3d^3(^4F)4s$ | $^5F_2$ | -6549 | -6344 | 11123 | 11503 | -1664 | -1023 |
| $3d^3(^4F)4s$ | $^5F_3$ | -6030 | -5829 | 11144 | 11523 | -1662 | -1021 |
| $3d^3(^4F)4s$ | $^5F_4$ | -5353 | -5161 | 11172 | 11550 | -1659 | -1018 |
| $3d^3(^4F)4s$ | $^5F_5$ | -4546 | -4370 | 11206 | 11583 | -1656 | -1015 |
| $3d^3(^4F)4s$ | $^3F_2$ | -6360 | -6151 | 10978 | 11294 | -1825 | -1186 |
| $3d^3(^4F)4s$ | $^3F_3$ | -5461 | -5264 | 11014 | 11329 | -1822 | -1183 |
| $3d^3(^4F)4s$ | $^3F_4$ | -4406 | -4230 | 11057 | 11369 | -1817 | -1179 |
| $3d^3(^4P)4s$ | $^5P_1$ | -6220 | -5930 | 11423 | 11099 | -1641 | -1022 |
| $3d^3(^4P)4s$ | $^5P_2$ | -6039 | -5733 | 11431 | 11106 | -1641 | -1021 |
| $3d^3(^4P)4s$ | $^5P_3$ | -5243 | -4975 | 11463 | 11135 | -1637 | -1018 |
| $3d^3(^2G)4s$ | $^3G_3$ | -5963 | -5750 | 11379 | 12046 | -1679 | -1046 |
| $3d^3(^2G)4s$ | $^3G_4$ | -5737 | -5510 | 11389 | 12062 | -1678 | -1046 |
| $3d^3(^2G)4s$ | $^3G_5$ | -5294 | -5060 | 11408 | 12090 | -1676 | -1043 |
| $3d^3(^4P)4s$ | $^3P_0$ | -6456 | -6095 | 11292 | 10915 | -1784 | -1168 |
| $3d^3(^4P)4s$ | $^3P_1$ | -6282 | -5961 | 11287 | 10907 | -1792 | -1174 |
| $3d^3(^2G)4s$ | $^1G_4$ | -5783 | -5496 | 11326 | 12007 | -1750 | -1118 |
| $3d^3(^4P)4s$ | $^3P_2$ | -5215 | -4886 | 11321 | 10930 | -1792 | -1178 |
| $3d^3(^2P)4s$ | $^3P_2$ | -5872 | -6026 | 11473 | 11169 | -1681 | -1053 |
| $3d^3(^2P)4s$ | $^3P_1$ | -5745 | -5776 | 11471 | 11151 | -1683 | -1058 |
| $3d^34s$ | $_0$ | -5292 | -5120 | 11470 | 11133 | -1693 | -1061 |
| $3d^3(^2D2)4s$ | $^3D_1$ | -6198 | -5455 | 11479 | 11230 | -1684 | -1056 |
| $3d^3(^2D2)4s$ | $^3D_3$ | -5043 | -4792 | 11549 | 11316 | -1664 | -1051 |
| $3d^3(^2D2)4s$ | $^3D_2$ | -4752 | -4162 | 11554 | 11288 | -1665 | -1047 |
| $3d^3(^2H)4s$ | $^3H_4$ | -5172 | -5102 | 11484 | 12628 | -1678 | -1031 |
| $3d^3(^2H)4s$ | $^3H_5$ | -5394 | -5254 | 11491 | 12676 | -1670 | -1025 |
| $3d^3(^2H)4s$ | $^3H_6$ | -5163 | -5008 | 11502 | 12696 | -1669 | -1023 |

TABLE III – *Continued from previous page*

| Level | | $q$ (cm$^{-1}$) | | $k_{\text{SMS}}$ (GHz·amu) | | $F$ (MHz·fm$^{-2}$) | |
|---|---|---|---|---|---|---|---|
| | | CI | CI +Σ | CI | CI +Σ | CI | CI +Σ |
| $3d^3(^2P)4s$ | $^1P_1$ | -4227 | -4171 | 11485 | 11107 | -1736 | -1117 |
| $3d^3(^2D2)4s$ | $^1D_2$ | -5030 | -4735 | 11472 | 11208 | -1745 | -1130 |
| $3d^3(^2H)4s$ | $^1H_5$ | -5216 | -5063 | 11423 | 12579 | -1750 | -1106 |
| $3d^3(^2F)4s$ | $^3F_4$ | -5711 | -5450 | 11768 | 11821 | -1650 | -1031 |
| $3d^3(^2F)4s$ | $^3F_3$ | -5545 | -5286 | 11775 | 11823 | -1650 | -1031 |
| $3d^3(^2F)4s$ | $^3F_2$ | -5392 | -5133 | 11782 | 11827 | -1649 | -1030 |
| $3d^3(^2F)4s$ | $^1F_3$ | -5426 | -5168 | 11703 | 11722 | -1731 | -1113 |
| $3d^3(^4F)4p$ | $^5G_2^o$ | -5411 | -5156 | 9556 | 10018 | -3673 | -3086 |
| $3d^3(^4F)4p$ | $^5G_3^o$ | -4710 | -4451 | 9594 | 10090 | -3672 | -3086 |
| $3d^3(^4F)4p$ | $^5G_4^o$ | -3812 | -3552 | 9643 | 10162 | -3671 | -3085 |
| $3d^3(^4F)4p$ | $^5G_5^o$ | -2728 | -2458 | 9704 | 10236 | -3669 | -3084 |
| $3d^3(^4F)4p$ | $^3D_1^o$ | -5473 | -5249 | 10207 | 10255 | -3672 | -3085 |
| $3d^3(^4F)4p$ | $^5D_2^o$ | -4706 | -4490 | 10389 | 10440 | -3671 | -3084 |
| $3d^3(^4F)4p$ | $^5G_6^o$ | -1431 | -1123 | 9778 | 10283 | -3669 | -3084 |
| $3d^3(^2D1)4s$ | $^3D_3$ | -5751 | -5410 | 12144 | 11467 | -1620 | -1002 |
| $3d^3(^4F)4p$ | $^5D_0^o$ | -3825 | -3480 | 10707 | 10573 | -3678 | -3091 |
| $3d^3(^4F)4p$ | $^3D_2^o$ | -3720 | -3321 | 10293 | 10545 | -3675 | -3091 |
| $3d^3(^2D1)4s$ | $^3D_2$ | -5566 | -5267 | 12153 | 11474 | -1620 | -1001 |
| $3d^3(^4F)4p$ | $^5D_3^o$ | -3810 | -3576 | 10542 | 10606 | -3669 | -3082 |
| $3d^3(^2D1)4s$ | $^3D_1$ | -5420 | -5159 | 12160 | 11479 | -1619 | -1001 |
| $3d^3(^4F)4p$ | $^5D_1^o$ | -3872 | -3478 | 10512 | 10571 | -3676 | -3091 |
| $3d^3(^4F)4p$ | $^5D_4^o$ | -2854 | -2616 | 10663 | 10688 | -3666 | -3080 |
| $3d^3(^4F)4p$ | $^5F_2^o$ | -3370 | -3046 | 10331 | 10398 | -3670 | -3085 |
| $3d^3(^4F)4p$ | $^5F_3^o$ | -1847 | -1568 | 10427 | 10536 | -3665 | -3079 |
| $3d^3(^4F)4p$ | $^5F_1^o$ | -3370 | -3046 | 10231 | 10465 | -3673 | -3089 |
| $3d^3(^4F)4p$ | $^5F_4^o$ | -2795 | -2501 | 10087 | 10312 | -3672 | -3089 |
| $3d^3(^4F)4p$ | $^5F_5^o$ | -2421 | -2095 | 10078 | 10398 | -3668 | -3086 |
| $3d^3(^4F)4p$ | $^3D_3^o$ | -3375 | -2997 | 10141 | 10425 | -3674 | -3091 |
| $3d^3(^2D1)4s$ | $^1D_2$ | -5464 | -5160 | 12079 | 11366 | -1702 | -1084 |
| $3d^3(^4F)4p$ | $^3G_3^o$ | -3737 | -3424 | 10482 | 10672 | -3673 | -3088 |
| $3d^3(^4F)4p$ | $^3G_4^o$ | -2984 | -2688 | 10505 | 10687 | -3670 | -3085 |
| $3d^3(^4F)4p$ | $^3G_5^o$ | -1993 | -1752 | 10531 | 10710 | -3665 | -3081 |
| $3d^3(^4F)4p$ | $^3F_2^o$ | -3673 | -3368 | 10457 | 10481 | -3670 | -3085 |
| $3d^3(^4F)4p$ | $^3F_3^o$ | -2860 | -2593 | 10492 | 10587 | -3667 | -3082 |
| $3d^3(^4F)4p$ | $^3F_4^o$ | -2111 | -1878 | 10529 | 10662 | -3664 | -3079 |
| $3d^3(^4P)4p$ | $^5P_1^o$ | -4269 | -3906 | 10678 | 10219 | -3651 | -3084 |
| $3d^3(^4P)4p$ | $^5P_2^o$ | -3630 | -3296 | 10708 | 10100 | -3650 | -3083 |
| $3d^3(^4P)4p$ | $^5D_0^o$ | -5187 | -4829 | 10079 | 9661 | -3649 | -3087 |
| $3d^3(^4P)4p$ | $^5P_3^o$ | -2798 | -2410 | 10767 | 10258 | -3651 | -3085 |
| $3d^3(^4P)4p$ | $^5D_1^o$ | -4829 | -4435 | 10088 | 9713 | -3648 | -3086 |
| $3d^3(^4P)4p$ | $^3P_2^o$ | -4498 | -4042 | 10195 | 9814 | -3648 | -3084 |
| $3d^3(^4P)4p$ | $^3H_4^o$ | -4650 | -4377 | 10022 | 10790 | -3647 | -3066 |
| $3d^3(^4P)4p$ | $^3P_0^o$ | -3587 | -3166 | 10265 | 9901 | -3652 | -3091 |
| $3d^3(^4P)4p$ | $^5D_2^o$ | -3157 | -2773 | 10196 | 9852 | -3650 | -3089 |
| $3d^3(^4P)4p$ | $^3P_1^o$ | -2874 | -2458 | 10362 | 9968 | -3650 | -3089 |
| $3d^3(^4P)4p$ | $^5D_3^o$ | -3443 | -2903 | 10087 | 9796 | -3650 | -3090 |
| $3d^3(^2G)4p$ | $^3H_5^o$ | -3914 | -3575 | 10109 | 11011 | -3647 | -3066 |
| $3d^3(^4P)4p$ | $^5D_4^o$ | -2140 | -1729 | 10088 | 9800 | -3650 | -3090 |
| $3d^3(^2G)4p$ | $^3H_6^o$ | -2252 | -1912 | 10141 | 11070 | -3647 | -3066 |
| $3d^3(^2G)4p$ | $^3G_3^o$ | -4149 | -4046 | 10703 | 11206 | -3649 | -3070 |
| $3d^3(^2G)4p$ | $^3G_4^o$ | -3153 | -3003 | 10788 | 11208 | -3649 | -3070 |
| $3d^3(^2G)4p$ | $^3G_5^o$ | -2653 | -2316 | 10773 | 11297 | -3649 | -3071 |
| $3d^3(^2G)4p$ | $^3F_4^o$ | -4260 | -3764 | 10744 | 11161 | -3649 | -3071 |
| $3d^3(^2P)4p$ | $^1S_0^o$ | -3329 | -2668 | 10230 | 9718 | -3642 | -3076 |
| $3d^3(^2G)4p$ | $^3F_2^o$ | -3363 | -3322 | 10757 | 10777 | -3650 | -3074 |
| $3d^3(^2G)4p$ | $^3F_3^o$ | -3258 | -2825 | 10789 | 11177 | -3650 | -3071 |
| $3d^3(^2P)4p$ | $^3P_1^o$ | -4531 | -4428 | 10691 | 10219 | -3645 | -3067 |
| $3d^3(^2G)4p$ | $^1G_4^o$ | -2774 | -2235 | 10826 | 11231 | -3649 | -3071 |
| $3d^3(^2P)4p$ | $^3P_0^o$ | -3280 | -3186 | 10639 | 10111 | -3646 | -3066 |

TABLE III – *Continued from previous page*TABLE III – *Continued from previous page*

| Level | | $q$ (cm$^{-1}$) | | $k_{\text{SMS}}$ (GHz·amu) | | $F$ (MHz·fm$^{-2}$) | |
|---|---|---|---|---|---|---|---|
| | | CI | CI +$\Sigma$ | CI | CI +$\Sigma$ | CI | CI +$\Sigma$ |
| $3d^3(^4P)4p$ | $^5S_2^o$ | -3665 | -2679 | 10177 | 9678 | -3648 | -3094 |
| $3d^3(^2G)4p$ | $^1F_3^o$ | -2437 | -2172 | 10615 | 10880 | -3644 | -3066 |
| $3d^3(^2P)4p$ | $^1D_2^o$ | -3840 | -3584 | 10378 | 10159 | -3644 | -3068 |
| $3d^3(^2G)4p$ | $^1H_5^o$ | -3195 | -2888 | 10262 | 11117 | -3643 | -3062 |
| $3d^3(^2P)4p$ | $^3P_2^o$ | -2610 | -2461 | 10618 | 10284 | -3644 | -3066 |
| $3d^3(^2P)4p$ | $^3D_1^o$ | -4422 | -4077 | 10457 | 10195 | -3647 | -3075 |
| $3d^3(^2H)4p$ | $^3H_4^o$ | -2742 | -2730 | 10782 | 11654 | -3643 | -3054 |
| $3d^3(^2H)4p$ | $^3H_5^o$ | -3231 | -3002 | 10745 | 11715 | -3643 | -3052 |
| $3d^3(^2P)4p$ | $^3D_2^o$ | -2437 | -2395 | 10474 | 10178 | -3643 | -3073 |
| $3d^3(^2H)4p$ | $^3H_6^o$ | -2872 | -2607 | 10758 | 11649 | -3643 | -3053 |
| $3d^3(^2P)4p$ | $^3D_3^o$ | -2726 | -2598 | 10682 | 10363 | -3647 | -3079 |
| $3d^3(^2D2)4p$ | $^1P_1^o$ | -3314 | -2768 | 11029 | 10329 | -3646 | -3079 |
| $3d^3(^2D2)4p$ | $^3F_2^o$ | -3090 | -3117 | 10699 | 10391 | -3638 | -3080 |
| $3d^3(^2P)4p$ | $^3S_1^o$ | -2574 | -2125 | 10995 | 10499 | -3648 | -3079 |
| $3d^3(^4P)4p$ | $^3D_3^o$ | -3231 | -2970 | 10960 | 10410 | -3648 | -3082 |
| $3d^3(^4P)4p$ | $^3D_1^o$ | -2595 | -2278 | 11074 | 10394 | -3643 | -3080 |
| $3d^3(^4P)4p$ | $^3D_2^o$ | -3374 | -2531 | 10902 | 10424 | -3646 | -3077 |
| $3d^3(^2D2)4p$ | $^3F_3^o$ | -3047 | -2456 | 10719 | 10501 | -3638 | -3075 |
| $3d^3(^2H)4p$ | $^3I_5^o$ | -3570 | -3381 | 10134 | 11549 | -3639 | -3050 |
| $3d^3(^2D2)4p$ | $^3F_4^o$ | -2748 | -1822 | 10658 | 10511 | -3641 | -3079 |
| $3d^3(^2H)4p$ | $^3I_6^o$ | -3085 | -2829 | 10141 | 11442 | -3640 | -3051 |
| $3d^3(^2D2)4p$ | $^3D_1^o$ | -4196 | -3628 | 10679 | 10331 | -3642 | -3077 |
| $3d^3(^2H)4p$ | $^3I_7^o$ | -2114 | -1822 | 10203 | 11539 | -3641 | -3052 |
| $3d^3(^2D2)4p$ | $^3D_2^o$ | -3023 | -2522 | 10714 | 10413 | -3640 | -3076 |
| $3d^3(^2H)4p$ | $^1G_4^o$ | -2648 | -2859 | 10698 | 11441 | -3638 | -3052 |
| $3d^3(^2D2)4p$ | $^3D_3^o$ | -2248 | -1844 | 10678 | 10308 | -3638 | -3075 |
| $3d^3(^2H)4p$ | $^1I_5^o$ | -3126 | -3000 | 10603 | 12229 | -3642 | -3049 |
| $3d^3(^2P)4p$ | $^3P_2^o$ | -2923 | -2469 | 10944 | 10394 | -3641 | -3084 |
| $3d^3(^2D2)4p$ | $^3P_1^o$ | -2868 | -2493 | 10982 | 10487 | -3642 | -3085 |
| $3d^3(^2D2)4p$ | $^3P_0^o$ | -2294 | -1941 | 10999 | 10342 | -3641 | -3084 |
| $3d^3(^2D2)4p$ | $^1F_3^o$ | -2144 | -2450 | 11207 | 11099 | -3639 | -3072 |
| $3d^3(^2H)4p$ | $^3G_5^o$ | -2762 | -2381 | 10984 | 11838 | -3638 | -3050 |
| $3d^3(^2H)4p$ | $^1I_6^o$ | -2744 | -2475 | 10676 | 11824 | -3640 | -3052 |
| $3d^3(^2H)4p$ | $^3G_4^o$ | -2425 | -2283 | 10995 | 11801 | -3640 | -3052 |
| $3d^3(^2H)4p$ | $^3G_3^o$ | -2951 | -1918 | 10984 | 11559 | -3639 | -3061 |
| $3d^3(^4P)4p$ | $^3S_1^o$ | -2637 | -2403 | 11122 | 10389 | -3642 | -3079 |
| $3d^3(^2D2)4p$ | $^1D_2^o$ | -2697 | -2374 | 11423 | 10799 | -3639 | -3076 |
| $3d^3(^2P)4p$ | $^1P_1^o$ | -1886 | -1598 | 10887 | 10410 | -3639 | -3070 |
| $3d^3(^2F)4p$ | $^3F_2^o$ | -3690 | -3323 | 10744 | 10673 | -3621 | -3060 |
| $3d^3(^2F)4p$ | $^3F_3^o$ | -3663 | -3279 | 10736 | 10661 | -3622 | -3060 |
| $3d^3(^2F)4p$ | $^3F_4^o$ | -3386 | -2941 | 10748 | 10729 | -3622 | -3061 |
| $3d^3(^2F)4p$ | $^3G_3^o$ | -3645 | -3315 | 10564 | 10815 | -3619 | -3055 |
| $3d^3(^2F)4p$ | $^3G_4^o$ | -3120 | -2732 | 10607 | 10700 | -3622 | -3057 |
| $3d^3(^2F)4p$ | $^3G_5^o$ | -2697 | -2311 | 10634 | 10755 | -3623 | -3060 |
| $3d^3(^2F)4p$ | $^3D_3^o$ | -3605 | -3233 | 11094 | 10842 | -3620 | -3053 |
| $3d^3(^2F)4p$ | $^1D_2^o$ | -2933 | -2595 | 11047 | 10629 | -3617 | -3048 |
| $3d^3(^2F)4p$ | $^3D_2^o$ | -2793 | -2413 | 11100 | 10793 | -3620 | -3051 |
| $3d^3(^2F)4p$ | $^3D_1^o$ | -2292 | -1918 | 11190 | 11012 | -3620 | -3053 |
| $3d^3(^2F)4p$ | $^1G_4^o$ | -2844 | -2538 | 10878 | 11015 | -3622 | -3053 |
| $3d^3(^2F)4p$ | $^1F_3^o$ | -2854 | -2512 | 11650 | 11315 | -3620 | -3054 |
| $3d^3(^2D1)4p$ | $^3D_1^o$ | -3654 | -3315 | 11374 | 10694 | -3588 | -3019 |
| $3d^3(^2D1)4p$ | $^3D_2^o$ | -3663 | -3298 | 11364 | 10477 | -3588 | -3019 |
| $3d^3(^2D1)4p$ | $^3D_3^o$ | -3278 | -2868 | 11386 | 10595 | -3589 | -3019 |
| $3d^3(^2D1)4p$ | $^1D_2^o$ | -3618 | -3364 | 11521 | 10639 | -3591 | -3028 |
| $3d^3(^2D1)4p$ | $^3F_2^o$ | -3275 | -2776 | 11077 | 10440 | -3587 | -3026 |
| $3d^3(^2D1)4p$ | $^3F_3^o$ | -3514 | -3036 | 11027 | 10470 | -3588 | -3026 |
| $3d^3(^2D1)4p$ | $^3F_4^o$ | -2778 | -2295 | 11065 | 10380 | -3591 | -3029 |
| $3d^3(^2D1)4p$ | $^3P_2^o$ | -3371 | -2910 | 11068 | 10257 | -3586 | -3026 |
| $3d^3(^2D1)4p$ | $^3P_1^o$ | -2685 | -2266 | 11114 | 10393 | -3587 | -3027 |

TABLE III – *Continued from previous page*

| Level | $q$ (cm$^{-1}$) | | $k_{\text{SMS}}$ (GHz·amu) | | $F$ (MHz·fm$^{-2}$) | |
|---|---|---|---|---|---|---|
| | CI | CI $+\Sigma$ | CI | CI $+\Sigma$ | CI | CI $+\Sigma$ |
| $3d^3(^2D1)4p\ ^3P_0^o$ | -2358 | -1949 | 11134 | 10182 | -3587 | -3027 |
| $3d^3(^2D1)4p\ ^1F_3^o$ | -2787 | -2418 | 11345 | 10596 | -3585 | -3019 |
| $3d^3(^2D1)4p\ ^1P_1^o$ | -3084 | -2675 | 11878 | 10818 | -3585 | -3021 |

TABLE IV: Calculated sensitivities to $\alpha$-variation $q$, and isotope shifts $\delta\nu_{A,A'}$ of selected transitions in Fe$^{4+}$. The asterisks (*) indicate transitions that may have underestimated errors due to differences in the state mixing between the CI and CI $+\Sigma$ calculations.

| $\lambda$ (nm) | Lower State | Upper State | $q$ (cm$^{-1}$) | $\delta\nu_{54,56}$ (ms$^{-1}$) | $\delta\nu_{57,56}$ (ms$^{-1}$) | $\delta\nu_{58,56}$ (ms$^{-1}$) |
|---|---|---|---|---|---|---|
| 119.6202 | $3d^3(^4F)4s\ ^3F_2$ | $3d^3(^2G)4p\ ^3G_3^o$ | 2105 (106) | 41.8 (15.2) | −27.3 (7.9) | −43.8 (14.2) |
| 119.6607 | $3d^3(^4F)4s\ ^3F_3$ | $3d^3(^2G)4p\ ^3G_4^o$ | 2261 (46) | 44.3 (9.1) | −28.5 (5.4) | −46.1 (8.4) |
| 120.1909 | $3d^3(^4F)4s\ ^3F_4$ | $3d^3(^2G)4p\ ^3G_5^o$ | 1914 (161) | 39.9 (17.3) | −26.5 (8.8) | −42 (16.1) |
| 123.4648 | $3d^3(^4P)4s\ ^3P_2$ | $3d^3(^4P)4p\ ^3S_1^o$ | 2482 (96) | 75.9 (28.1) | −43.8 (13.8) | −75.8 (26.2) |
| 125.8791 | $3d^3(^4F)4s\ ^3F_3$ | $3d^3(^4P)4p\ ^3P_2^o$ | 1223 (260) | 156.3 (58.0) | −82.2 (27.8) | −150.9 (54.1) |
| 128.0471 | $3d^3(^2G)4s\ ^3G_5$ | $3d^3(^2D2)4p\ ^3F_4^o$ | 3238 (691) | 156.9 (70.2) | −83.2 (33.6) | −151.9 (65.5) |
| 128.4109 | $3d^3(^4P)4s\ ^5P_1$ | $3d^3(^4P)4p\ ^5S_2^o*$ | 3251 (696) | 142.1 (15.5) | −76.4 (8.3) | −138.3 (14.4) |
| 128.5918 | $3d^3(^4P)4s\ ^3P_1$ | $3d^3(^2D2)4p\ ^3P_1^o$ | 3468 (53) | 63.8 (10.6) | −38.5 (6.2) | −64.7 (9.8) |
| 128.7046 | $3d^3(^2G)4s\ ^1G_4$ | $3d^3(^2D2)4p\ ^1F_3^o*$ | 3046 (593) | 103.6 (67.3) | −57.5 (32.1) | −102 (62.7) |
| 128.8169 | $3d^3(^4P)4s\ ^5P_2$ | $3d^3(^4P)4p\ ^5S_2^o*$ | 3054 (680) | 142.8 (15.5) | −76.7 (8.3) | −138.9 (14.4) |
| 129.1187 | $3d^3(^4P)4s\ ^5P_2$ | | | | | |
| 129.3377 | $3d^3(^2G)4s\ ^3G_3$ | $3d^3(^2D2)4p\ ^3F_2^o$ | 2634 (239) | 164 (83.5) | −86.7 (39.8) | −158.6 (77.8) |
| 129.7547 | $3d^3(^4P)4s\ ^5P_3$ | $3d^3(^4P)4p\ ^5S_2^o*$ | 2296 (718) | 145.4 (15.3) | −78 (8.3) | −141.3 (14.2) |
| 130.0608 | $3d^3(^4P)4s\ ^5P_3$ | $3d^3(^2P)4p\ ^1D_2^o$ | 1391 (11) | 105.2 (10.4) | −58.8 (6.3) | −103.9 (9.6) |
| 130.2787 | $3d^3(^4P)4s\ ^3P_2$ | $3d^3(^2P)4p\ ^3P_2^o$ | 2425 (179) | 83.7 (6.5) | −47.9 (4.7) | −83.3 (5.9) |
| 131.1828 | $3d^3(^2P)4s\ ^1P_1$ | $3d^3(^2D2)4p\ ^1D_2^o$ | 1797 (267) | 51.2 (21.8) | −32.9 (11.0) | −53.3 (20.3) |
| 131.4492 | $3d^3(^2D1)4s\ ^3D_3$ | $3d^3(^2D1)4p\ ^3P_2^o$ | 2500 (119) | 126.6 (12.4) | −69 (7.1) | −123.8 (11.5) |
| 132.0410 | $3d^3(^2H)4s\ ^3H_4$ | $3d^3(^2H)4p\ ^3G_3^o*$ | 3183 (962) | 114.2 (49.9) | −63.2 (24) | −112.3 (46.5) |
| 132.1341 | $3d^3(^2G)4s\ ^3G_4$ | $3d^3(^2H)4p\ ^3H_5^o$ | 2508 (2) | 52.2 (26.2) | −33.6 (13.0) | −54.3 (24.4) |
| 132.1490 | $3d^3(^2G)4s\ ^3G_5$ | $3d^3(^2H)4p\ ^3H_6^o$ | 2453 (31) | 60.1 (18.8) | −37.4 (9.7) | −61.8 (17.5) |
| 132.1850 | $3d^3(^2H)4s\ ^3H_4$ | $3d^3(^2H)4p\ ^3G_4^o$ | 2819 (72) | 93.4 (29.9) | −53.3 (14.7) | −92.9 (27.8) |
| 132.3269 | $3d^3(^2H)4s\ ^3H_5$ | $3d^3(^2H)4p\ ^3G_4^o$ | 2971 (1) | 97.4 (33.5) | −55.2 (16.4) | −96.6 (31.2) |
| 132.5781 | $3d^3(^2H)4s\ ^3H_5$ | $3d^3(^2H)4p\ ^3G_5^o$ | 2874 (241) | 94.2 (29.4) | −53.7 (14.5) | −93.6 (27.4) |
| 132.7101 | $3d^3(^2D2)4s\ ^1D_2$ | $3d^3(^2P)4p\ ^1P_1^o$ | 3137 (7) | 94.4 (19.2) | −53.4 (9.9) | −93.5 (17.9) |
| 133.0401 | $3d^3(^2H)4s\ ^3H_6$ | $3d^3(^2H)4p\ ^3G_5^o$ | 2627 (226) | 95.8 (30.3) | −54.5 (14.9) | −95.2 (28.2) |
| 133.1185 | $3d^3(^2D2)4s\ ^3D_1$ | $3d^3(^2D2)4p\ ^3P_0^o$ | 3514 (389) | 98.4 (36.2) | −55.7 (17.6) | −97.6 (33.7) |
| 133.1640 | $3d^3(^2D2)4s\ ^1D_2$ | $3d^3(^2D2)4p\ ^1D_2^o$ | 2361 (28) | 59.8 (32.0) | −37 (15.7) | −61.3 (29.8) |
| 135.0535 | $3d^3(^4P)4s\ ^3P_1$ | $3d^3(^4P)4p\ ^3D_2^o*$ | 3430 (522) | 67.5 (9.7) | −40.6 (6.0) | −68.5 (9.0) |
| 135.1755 | $3d^3(^4F)4s\ ^5F_1$ | $3d^3(^4F)4p\ ^3D_2^o$ | 3369 (192) | 101.4 (12.3) | −57.5 (7.1) | −100.6 (11.4) |
| 135.3737 | $3d^3(^2D2)4s\ ^3D_3$ | $3d^3(^2P)4p\ ^3P_2^o$ | 2332 (100) | 111.6 (10.1) | −62.1 (6.3) | −110 (9.3) |
| 135.4847 | $3d^3(^2F)4s\ ^1F_3$ | $3d^3(^2F)4p\ ^1F_3^o$ | 2656 (83) | 59 (32.0) | −36.8 (15.7) | −60.6 (29.8) |
| 135.7675 | $3d^3(^2F)4s\ ^3F_4$ | $3d^3(^2F)4p\ ^3D_3^o$ | 2216 (110) | 106.7 (27.8) | −59.8 (13.8) | −105.4 (25.9) |
| 135.8567 | $3d^3(^2D2)4s\ ^3D_1$ | $3d^3(^2D2)4p\ ^3D_2^o$ | 2933 (241) | 92.2 (6.7) | −52.9 (5.1) | −91.9 (6.1) |
| 135.9006 | $3d^3(^2D1)4s\ ^3D_3$ | $3d^3(^2D1)4p\ ^3F_4^o$ | 3115 (142) | 116.1 (4.8) | −64.3 (4.6) | −114.1 (4.2) |
| 136.1447 | $3d^3(^2F)4s\ ^1F_3$ | $3d^3(^2F)4p\ ^1G_4^o$ | 2630 (47) | 85.8 (11.5) | −49.5 (6.7) | −85.6 (10.7) |
| 136.1825 | $3d^3(^2F)4s\ ^3F_4$ | $3d^3(^2F)4p\ ^3G_5^o$ | 3138 (124) | 114.2 (7.7) | −63.4 (5.5) | −112.4 (7.1) |
| 136.2864 | $3d^3(^2D2)4s\ ^3D_3$ | $3d^3(^2D2)4p\ ^3D_3^o$ | 2948 (153) | 109.1 (13.1) | −61 (7.4) | −107.7 (12.2) |
| 136.3077 | $3d^3(^4F)4s\ ^5F_3$ | $3d^3(^4F)4p\ ^5F_4^o$ | 3279 (157) | 125.5 (14.7) | −69 (8.1) | −123.1 (13.6) |
| 136.3642 | $3d^3(^4F)4s\ ^5F_4$ | $3d^3(^4F)4p\ ^5F_5^o$ | 3066 (133) | 120.3 (7.2) | −66.5 (5.4) | −118.2 (6.6) |
| 136.4824 | $3d^3(^4F)4s\ ^5F_2$ | $3d^3(^4F)4p\ ^3D_1^o$ | 1095 (18) | 129.1 (30.3) | −70.7 (15.0) | −126.4 (28.2) |
| 136.4984 | $3d^3(^2F)4s\ ^3F_2$ | $3d^3(^2F)4p\ ^3D_3^o$ | 1900 (112) | 107.2 (27.3) | −60.1 (13.6) | −105.9 (25.4) |
| 136.5115 | $3d^3(^4F)4s\ ^5F_3$ | $3d^3(^4F)4p\ ^3D_2^o$ | 2508 (197) | 104.4 (12.4) | −59 (7.2) | −103.4 (11.5) |
| 136.5571 | $3d^3(^4F)4s\ ^5F_2$ | $3d^3(^4F)4p\ ^5F_3^o$ | 4776 (74) | 104 (24.9) | −58.7 (12.5) | −103 (23.2) |
| 137.0303 | $3d^3(^2H)4s\ ^3H_6$ | $3d^3(^2H)4p\ ^1H_5^o$ | 2008 (28) | 60 (39.4) | −37.7 (19.1) | −61.9 (36.7) |
| 137.0947 | $3d^3(^4F)4s\ ^5F_1$ | $3d^3(^4F)4p\ ^5F_2^o$ | 4189 (87) | 114.9 (28.7) | −64 (14.2) | −113.3 (26.7) |
| 137.1987 | $3d^3(^2D2)4s\ ^3D_1$ | $3d^3(^2D2)4p\ ^3D_1^o$ | 1827 (175) | 99.4 (10.2) | −56.4 (6.3) | −98.7 (9.4) |
| 137.3587 | $3d^3(^4F)4s\ ^5F_4$ | $3d^3(^4F)4p\ ^5F_4^o$ | 2611 (166) | 127.9 (14.7) | −70.2 (8.1) | −125.4 (13.6) |
| 137.3967 | $3d^3(^4P)4s\ ^3P_2$ | $3d^3(^4P)4p\ ^3D_3^o$ | 1916 (68) | 70.2 (15.1) | −42 (8.1) | −71 (14.0) |

TABLE IV – *Continued from previous page*

| λ (nm) | Lower State | Upper State | q (cm$^{-1}$) | $\delta\nu_{54,56}$ (ms$^{-1}$) | $\delta\nu_{57,56}$ (ms$^{-1}$) | $\delta\nu_{58,56}$ (ms$^{-1}$) |
|---|---|---|---|---|---|---|
| 137.4116 | $3d^3(^2F)4s$ $^3F_3$ | $3d^3(^2F)4p$ $^3G_4^o$ | 2554 (128) | 119.5 (6.3) | −66 (5.1) | −117.4 (5.7) |
| 137.4789 | $3d^3(^2D1)4s$ $^3D_2$ | $3d^3(^2D1)4p$ $^3F_3^o$ | 2231 (179) | 108.8 (12.0) | −60.9 (7.0) | −107.4 (11.1) |
| 137.6337 | $3d^3(^4F)4s$ $^5F_5$ | $3d^3(^4F)4p$ $^5F_5^o$ | 2275 (149) | 123.2 (7.2) | −68 (5.4) | −121 (6.6) |
| 137.6455 | $3d^3(^4F)4s$ $^5F_2$ | $3d^3(^4F)4p$ $^5F_2^o$ | 3843 (88) | 116.2 (28.8) | −64.6 (14.3) | −114.4 (26.8) |
| 137.8560 | $3d^3(^4P)4s$ $^5P_3$ | $3d^3(^4P)4p$ $^5D_4^o$ | 3246 (143) | 136.8 (6.1) | −74.5 (5.1) | −133.7 (5.5) |
| 137.9040 | $3d^3(^2G)4s$ $^3G_5$ | $3d^3(^2G)4p$ $^1G_4^o$ | 2825 (304) | 95.4 (25.6) | −54.6 (12.8) | −94.9 (23.9) |
| 138.0112 | $3d^3(^4F)4s$ $^5F_1$ | $3d^3(^4F)4p$ $^5D_1^o$ | 3212 (187) | 99 (29.6) | −56.5 (14.7) | −98.4 (27.6) |
| 138.4055 | $3d^3(^2P)4s$ $^3P_1$ | $3d^3(^4P)4p$ $^3D_2^o$ * | 3245 (874) | 83.7 (15.2) | −49 (8.3) | −84 (14.1) |
| 138.5313 | $3d^3(^4F)4s$ $^5F_1$ | $3d^3(^4F)4p$ $^5D_0^o$ | 3209 (138) | 98.7 (47.2) | −56.4 (22.8) | −98.2 (44.0) |
| 138.7938 | $3d^3(^2H)4s$ $^3H_6$ | $3d^3(^2H)4p$ $^3I_7^o$ | 3186 (136) | 122.6 (13.9) | −67.6 (7.8) | −120.4 (12.9) |
| 138.8195 | $3d^3(^4P)4s$ $^5P_1$ | $3d^3(^4P)4p$ $^3P_1^o$ | 3472 (126) | 118.5 (8.1) | −65.8 (5.7) | −116.7 (7.4) |
| 138.8328 | $3d^3(^4P)4s$ $^5P_1$ | $3d^3(^4P)4p$ $^5D_2^o$ | 3157 (94) | 129.2 (5.3) | −70.9 (4.9) | −126.7 (4.7) |
| 139.3073 | $3d^3(^4P)4s$ $^5P_2$ | $3d^3(^4P)4p$ $^5D_2^o$ | 2960 (78) | 129.9 (5.3) | −71.3 (4.9) | −127.3 (4.7) |
| 139.4272 | $3d^3(^2G)4s$ $^3G_4$ | $3d^3(^2G)4p$ $^3F_3^o$ | 2685 (205) | 97.8 (26.8) | −55.8 (13.4) | −97.3 (24.9) |
| 139.4665 | $3d^3(^2G)4s$ $^3G_3$ | $3d^3(^2G)4p$ $^3F_2^o$ | 2428 (171) | 133.1 (59.8) | −72.6 (28.7) | −130.2 (55.8) |
| 139.7106 | $3d^3(^2P)4s$ $^3P_1$ | $3d^3(^2P)4p$ $^3S_1^o$ | 3650 (479) | 76.4 (13.8) | −45.7 (6.8) | −77.4 (13.0) |
| 139.7972 | $3d^3(^4P)4s$ $^5P_3$ | $3d^3(^4P)4p$ $^5D_3^o$ | 2072 (272) | 137.5 (6.1) | −74.9 (5.1) | −134.4 (5.5) |
| 140.0243 | $3d^3(^4F)4s$ $^3F_2$ | $3d^3(^4F)4p$ $^3F_2^o$ | 2782 (95) | 96.8 (27.4) | −54.8 (13.6) | −96 (25.5) |
| 140.3370$^a$ | $3d^3(^2G)4s$ $^3G_4$ | $3d^3(^2G)4p$ $^3F_4^o$ | 1745 (268) | 99.2 (21.3) | −56.6 (10.4) | −98.7 (20.1) |
| 140.4260 | $3d^3(^4P)4s$ $^3P_0$ | $3d^3(^2P)4p$ $^3D_1^o$ | 2018 (16) | 87.8 (11.6) | −50.6 (6.8) | −87.6 (10.7) |
| 140.6669 | $3d^3(^4F)4s$ $^3F_4$ | $3d^3(^4F)4p$ $^3F_4^o$ | 2352 (57) | 86.9 (17.2) | −50.2 (9.1) | −86.8 (16) |
| 140.7248 | $3d^3(^2H)4s$ $^1H_5$ | $3d^3(^2H)4p$ $^1I_6^o$ | 2587 (115) | 89.2 (4.8) | −51.5 (4.6) | −89.1 (4.2) |
| 140.8117 | $3d^3(^4F)4s$ $^5F_1$ | $3d^3(^4F)4p$ $^5F_2^o$ | 3299 (120) | 110 (14.4) | −61.9 (8.1) | −108.8 (13.4) |
| 140.9026 | $3d^3(^4F)4s$ $^5F_3$ | $3d^3(^4F)4p$ $^5D_2^o$ | 1339 (15) | 114.4 (31.0) | −64 (15.3) | −112.9 (28.8) |
| 140.9220 | $3d^3(^4F)4s$ $^5F_4$ | $3d^3(^4F)4p$ $^5D_3^o$ | 1585 (42) | 101.4 (29.7) | −57.8 (14.7) | −100.8 (27.6) |
| 141.6219 | $3d^3(^2D2)4s$ $^1D_2$ | $3d^3(^2D2)4p$ $^1F_3^o$ * | 2285 (601) | 28.8 (15.3) | −22.8 (8.3) | −32.8 (14.2) |
| 142.0602 | $3d^3(^2G)4s$ $^3G_4$ | $3d^3(^2G)4p$ $^3G_4^o$ | 2506 (77) | 94.7 (24.3) | −54.6 (12.3) | −94.5 (22.7) |
| 142.2481 | $3d^3(^2P)4s$ $^3P_1$ | $3d^3(^2P)4p$ $^3D_2^o$ | 3381 (73) | 106.4 (5.4) | −60 (4.9) | −105.3 (4.8) |
| 142.9004 | $3d^3(^2G)4s$ $^3G_5$ | $3d^3(^2G)4p$ $^3G_4^o$ | 2057 (83) | 97.1 (25.2) | −55.8 (12.7) | −96.8 (23.5) |
| 143.0309 | $3d^3(^2F)4s$ $^1F_3$ | $3d^3(^2F)4p$ $^1D_2^o$ | 2573 (79) | 121.4 (41.6) | −66.9 (20.2) | −119.2 (38.8) |
| 143.0573 | $3d^3(^4F)4s$ $^5F_5$ | $3d^3(^4F)4p$ $^5G_6^o$ | 3247 (131) | 134.7 (13.0) | −73.8 (7.6) | −132 (12.1) |
| 143.0751 | $3d^3(^2D2)4s$ $^3D_3$ | $3d^3(^4P)4p$ $^3D_3^o$ | 1823 (10) | 99.2 (30.4) | −56.8 (15.0) | −98.7 (28.3) |
| 144.0528 | $3d^3(^4F)4s$ $^5F_4$ | $3d^3(^4F)4p$ $^5G_5^o$ | 2704 (78) | 136.4 (15.5) | −74.7 (8.5) | −133.6 (14.3) |
| 144.0792 | $3d^3(^4P)4s$ $^5P_1$ | $3d^3(^4P)4p$ $^5P_2^o$ | 2635 (45) | 106.6 (27.5) | −60.5 (13.8) | −105.8 (25.6) |
| 144.1049 | $3d^3(^4P)4s$ $^5P_3$ | $3d^3(^4P)4p$ $^5P_3^o$ | 2565 (120) | 94.7 (18.0) | −54.9 (9.6) | −94.7 (16.7) |
| 144.2221 | $3d^3(^2G)4s$ $^1G_4$ | $3d^3(^2G)4p$ $^1H_5^o$ | 2608 (20) | 101.7 (17.2) | −57.6 (9.2) | −100.9 (16.0) |
| 144.6618$^a$ | $3d^3(^2G)4s$ $^3G_5$ | $3d^3(^2G)4p$ $^3H_6^o$ | 3148 (106) | 110.4 (24.2) | −62.1 (12.3) | −109.2 (22.5) |
| 144.8494 | $3d^3(^2G)4s$ $^1G_4$ | $3d^3(^2G)4p$ $^1F_3^o$ | 3324 (21) | 124.1 (40.1) | −68.3 (19.5) | −121.8 (37.3) |
| 144.8846 | $3d^3(^4F)4s$ $^5F_3$ | $3d^3(^4F)4p$ $^5G_4^o$ | 2277 (59) | 141.2 (14.3) | −77 (8.1) | −138.1 (13.3) |
| 144.9928 | $3d^3(^2F)4s$ $^3F_4$ | $3d^3(^2F)4p$ $^3F_4^o$ | 2508 (183) | 117 (8.5) | −65.4 (5.9) | −115.4 (7.8) |
| 145.3618 | $3d^3(^2H)4s$ $^1H_5$ | $3d^3(^2H)4p$ $^1H_5^o$ | 2063 (26) | 49.8 (45.4) | −33 (21.9) | −52.5 (42.3) |
| 145.6161 | $3d^3(^4F)4s$ $^5F_2$ | $3d^3(^4F)4p$ $^5G_3^o$ | 1893 (54) | 146.3 (12.3) | −79.5 (7.3) | −142.9 (11.4) |
| 145.6285 | $3d^3(^4P)4s$ $^5P_2$ | $3d^3(^4P)4p$ $^5P_1^o$ | 1828 (58) | 95.7 (13.9) | −55.4 (7.9) | −95.7 (12.9) |
| 145.7727 | $3d^3(^4P)4s$ $^5P_3$ | $3d^3(^4P)4p$ $^5P_2^o$ | 1679 (66) | 109.8 (27.5) | −62.2 (13.8) | −108.9 (25.6) |
| 146.0726 | $3d^3(^4F)4s$ $^5F_4$ | $3d^3(^4F)4p$ $^5G_4^o$ | 1609 (68) | 144 (14.5) | −78.4 (8.2) | −140.7 (13.4) |
| 146.2631 | $3d^3(^4F)4s$ $^5F_1$ | $3d^3(^4F)4p$ $^5G_2^o$ | 1534 (48) | 152 (9.5) | −82.2 (6.3) | −148.3 (8.7) |
| 146.4876 | $3d^3(^2F)4s$ $^3F_2$ | $3d^3(^2F)4p$ $^3F_2^o$ | 1810 (108) | 123.2 (12.4) | −68.4 (7.3) | −121.2 (11.5) |
| 146.6649 | $3d^3(^2G)4s$ $^3G_4$ | $3d^3(^2G)4p$ $^3H_5^o$ | 1935 (112) | 113.6 (22.7) | −63.8 (11.6) | −112.3 (21.1) |
| 146.8911 | $3d^3(^4F)4s$ $^5F_2$ | $3d^3(^4F)4p$ $^5G_2^o$ | 1188 (49) | 153.5 (9.6) | −83 (6.4) | −149.7 (8.8) |
| 146.9000 | $3d^3(^2H)4s$ $^3H_6$ | $3d^3(^2H)4p$ $^3H_6^o$ | 2401 (110) | 112.7 (29.8) | −63.4 (14.8) | −111.5 (27.8) |
| 147.2098 | $3d^3(^2H)4s$ $^3H_5$ | $3d^3(^2H)4p$ $^3H_5^o$ | 2252 (89) | 104.5 (21.6) | −59.6 (11.1) | −103.9 (20.1) |
| 147.2512 | $3d^3(^2H)4s$ $^3H_4$ | $3d^3(^2H)4p$ $^3H_4^o$ | 2372 (57) | 106 (27.0) | −60.2 (13.5) | −105.2 (25.1) |
| 147.5604 | $3d^3(^2G)4s$ $^3G_5$ | $3d^3(^2G)4p$ $^3H_5^o$ | 1485 (105) | 116.1 (22.1) | −65.1 (11.4) | −114.7 (20.6) |
| 147.9471 | $3d^3(^2G)4s$ $^3G_4$ | $3d^3(^2G)4p$ $^3H_4^o$ | 1132 (44) | 135.2 (10.5) | −74.2 (6.7) | −132.5 (9.7) |
| 149.6266 | $3d^3(^2G)4s$ $^1G_4$ | $3d^3(^2G)4p$ $^3F_4^o$ | 1732 (209) | 96.7 (26.6) | −55.6 (13.3) | −96.4 (24.7) |
| 153.0439 | $3d^3(^2D2)4s$ $^1D_2$ | $3d^3(^2D2)4p$ $^1P_1^o$ | 1967 (251) | 100 (44.4) | −57.4 (21.5) | −99.6 (41.4) |
| 154.3234 | $3d^3(^4F)4s$ $^3F_2$ | $3d^3(^4F)4p$ $^3D_1^o$ | 901 (13) | 118.7 (27.9) | −66.1 (13.9) | −117 (25.9) |
| 155.4219 | $3d^3(^4F)4s$ $^3F_4$ | $3d^3(^4F)4p$ $^3D_3^o$ | 1234 (202) | 108.3 (5.8) | −61.3 (5.2) | −107.4 (5.2) |

$^a$ In [31], these lines are unresolved and could be misidentified.

TABLE IV – *Continued from previous page*

| $\lambda$ (nm) | Lower State | Upper State | q (cm$^{-1}$) | $\delta\nu_{54,56}$ (ms$^{-1}$) | $\delta\nu_{57,56}$ (ms$^{-1}$) | $\delta\nu_{58,56}$ (ms$^{-1}$) |
|---|---|---|---|---|---|---|

TABLE V: Calculated energies and $q$-values (relative to the ground state $3d^6$ $^5D_4$) and Landé $g$-factors of $4s$ and $4p$ of energy levels in Ni$^{4+}$.

| Level | | Energy | | $g$-factor | | q (cm$^{-1}$) |
|---|---|---|---|---|---|---|
| | | CI +$\Sigma$ | Exp. [25] | CI +$\Sigma$ | LS | |
| $3d^5(^6S)4s$ | $^7S_3$ | 161376 | 164526 | 1.9991 | 2 | -6960 |
| $3d^5(^6S)4s$ | $^5S_2$ | 175638 | 178020 | 1.9985 | 2 | -6447 |
| $3d^5(^4G)4s$ | $^5G_6$ | 206960 | 208046 | 1.3329 | 1.333 | -7062 |
| $3d^5(^4G)4s$ | $^5G_5$ | 206965 | 208131 | 1.2665 | 1.2666 | -7031 |
| $3d^5(^4G)4s$ | $^5G_2$ | 206820 | 208152 | 0.3360 | 0.3333 | -7299 |
| $3d^5(^4G)4s$ | $^5G_4$ | 206930 | 208164 | 1.1503 | 1.15 | -7091 |
| $3d^5(^4G)4s$ | $^5G_3$ | 206873 | 208165 | 0.9178 | 0.9166 | -7195 |
| $3d^5(^4P)4s$ | $^5P_3$ | 212267 | 212096 | 1.6411 | 1.6667 | -8160 |
| $3d^5(^4P)4s$ | $^5P_2$ | 212451 | 212253 | 1.7889 | 1.8333 | -7860 |
| $3d^5(^4P)4s$ | $^5P_1$ | 212727 | 212456 | 2.4308 | 2.5 | -7393 |
| $3d^5(^4D)4s$ | $^5D_4$ | 215975 | 216190 | 1.4871 | 1.5 | -7159 |
| $3d^5(^4D)4s$ | $^5D_1$ | 216201 | 216435 | 1.5650 | 1.5 | -6679 |
| $3d^5(^4D)4s$ | $^5D_2$ | 216469 | 216591 | 1.5398 | 1.5 | -6265 |
| $3d^5(^4D)4s$ | $^5D_3$ | 216518 | 216596 | 1.5098 | 1.5 | -6221 |
| $3d^5(^4G)4s$ | $^3G_5$ | 216442 | 217049 | 1.1994 | 1.2 | -6723 |
| $3d^5(^4G)4s$ | $^3G_3$ | 216298 | 217101 | 0.7626 | 0.75 | -6962 |
| $3d^5(^4G)4s$ | $^3G_4$ | 216431 | 217129 | 1.0621 | 1.05 | -6709 |
| $3d^5(^4P)4s$ | $^3P_2$ | 221778 | 221088 | 1.4475 | 1.5 | -7849 |
| $3d^5(^4P)4s$ | $^3P_1$ | 222149 | 222429 | 1.4014 | 1.5 | -7136 |
| $3d^5(^4D)4s$ | $^3D_3$ | 225486 | 225201 | 1.3314 | 1.3333 | -6770 |
| $3d^5(^4D)4s$ | $^3D_1$ | 225865 | 225545 | 0.5927 | 0.5 | -6044 |
| $3d^5(^4D)4s$ | $^3D_2$ | 226048 | 225617 | 1.2088 | 1.1667 | -5846 |
| $3d^5(^2I)4s$ | $^3I_6$ | 228688 | 229409 | 1.0252 | 1.0238 | -7225 |
| $3d^5(^2I)4s$ | $^3I_5$ | 228624 | 229413 | 0.8357 | 0.8333 | -7340 |
| $3d^5(^2I)4s$ | $^3I_7$ | 228804 | 229441 | 1.1429 | 1.1428 | -6984 |
| $3d^5(^2D3)4s$ | $^3D_3$ | 233464 | 232546 | 1.2718 | 1.333 | -8403 |
| $3d^5(^2D3)4s$ | $^3D_2$ | 233636 | 232656 | 1.0541 | 1.1667 | -8239 |
| $3d^5(^2D3)4s$ | $^3D_1$ | 233948 | 232911 | 0.3490 | 0.5 | -7838 |
| $3d^5(^2I)4s$ | $^1I_6$ | 233405 | 233839 | 1.0027 | 1 | -7078 |
| $3d^5(^4F)4s$ | $^5F_5$ | 235539 | 234082 | 1.3931 | 1.4 | -7340 |
| $3d^5(^4F)4s$ | $^5F_4$ | 235474 | 234125 | 1.3328 | 1.35 | -7495 |
| $3d^5(^4F)4s$ | $^5F_3$ | 235590 | 234275 | 1.2302 | 1.25 | -7232 |
| $3d^5(^4F)4s$ | $^5F_2$ | 235649 | 234413 | 0.9136 | 1 | -7353 |
| $3d^5(^2F1)4s$ | $^3F_4$ | 236638 | 235421 | 1.2612 | 1.25 | -6580 |
| $3d^5(^2F1)4s$ | $^3F_2$ | 236988 | 235737 | 0.9453 | 1 | -6269 |
| $3d^5(^4F)4s$ | $^5F_1$ | 236450 | 236117 | 0.1604 | 0 | -5828 |
| $3d^5(^2F1)4s$ | $^3F_3$ | 237751 | 236454 | 1.1382 | 1.08333 | -5420 |
| $3d^5(^2D3)4s$ | $^1D_2$ | 240573 | 239108 | 0.9180 | 0.6667 | -4743 |
| $3d^5(^2F1)4s$ | $^1F_3$ | 241603 | 240193 | 0.9904 | 1 | -6201 |
| $3d^5(^2H)4s$ | $^3H_4$ | 241859 | 240960 | 0.8957 | 0.8 | -7936 |
| $3d^5(^2H)4s$ | $^3H_5$ | 242009 | 241082 | 1.0978 | 1.0333 | -7590 |
| $3d^5(^2H)4s$ | $^3H_6$ | 242951 | 241774 | 1.1631 | 1.16667 | -6291 |
| $3d^5(^2G2)4s$ | $^3G_3$ | 243251 | 242290 | 0.8036 | 0.75 | -6434 |
| $3d^5(^2G2)4s$ | $^3G_4$ | 243603 | 242504 | 0.9836 | 1.05 | -6387 |
| $3d^5(^6S)4p$ | $^7P_2^o$ | 240905 | 242837 | 2.3276 | 2.3333 | -5019 |
| $3d^5(^2G2)4s$ | $^3G_5$ | 243904 | 242863 | 1.1335 | 1.2 | -5864 |
| $3d^5(^4F)4s$ | $^3F_2$ | 245147 | 243266 | 0.6806 | 0.6667 | -6547 |
| $3d^5(^4F)4s$ | $^3F_4$ | 245028 | 243332 | 1.2041 | 1.25 | -6892 |
| $3d^5(^4F)4s$ | $^3F_3$ | 245219 | 243371 | 1.0676 | 1.0833 | -6517 |
| $3d^5(^6S)4p$ | $^7P_3^o$ | 241735 | 243609 | 1.9116 | 1.9166 | -4272 |
| $3d^5(^6S)4p$ | $^7P_4^o$ | 243090 | 244901 | 1.7493 | 1.75 | -2726 |
| $3d^5(^2H)4s$ | $^1H_5$ | 247646 | 246241 | 1.0072 | 1 | -6190 |
| $3d^5(^2G2)4s$ | $^1G_4$ | 248295 | 247049 | 1.0600 | 1 | -6125 |
| $3d^5(^2F2)4s$ | $^3F_3$ | 248495 | 247105 | 1.0845 | 1.08333 | -6791 |

TABLE V – *Continued from previous page*

| Level | | Energy | | g-factor | | q (cm$^{-1}$) |
|---|---|---|---|---|---|---|
| | | CI +Σ | Exp. [25] | CI +Σ | LS | |
| $3d^5(^2F2)4s$ | $^3F_2$ | 248504 | 247165 | 0.6704 | 0.6667 | -6690 |
| $3d^5(^2F2)4s$ | $^3F_4$ | 248806 | 247282 | 1.2127 | 1.25 | -6167 |
| $3d^5(^2F2)4s$ | $^1F_3$ | 253353 | 251655 | 1.0033 | 1 | -6483 |
| $3d^5(^6S)4p$ | $^5P_3^o$ | 252631 | 253863 | 1.6699 | 1.666 | -3726 |
| $3d^5(^2S)4s$ | $^3S_1$ | 256093 | 253905 | 1.9994 | 2 | -6879 |
| $3d^5(^6S)4p$ | $^5P_2^o$ | 253294 | 254496 | 1.8365 | 1.833 | -3143 |
| $3d^5(^6S)4p$ | $^5P_1^o$ | 253703 | 254885 | 2.4978 | 2.5 | -2796 |
| $3d^5(^2D2)4s$ | $^3D_1$ | 266973 | 263701 | 0.5006 | 0.5 | -6959 |
| $3d^5(^2D2)4s$ | $^3D_2$ | 267037 | 263736 | 1.1657 | 1.1667 | -6841 |
| $3d^5(^2D2)4s$ | $^3D_3$ | 267155 | 263806 | 1.3310 | 1.333 | -6614 |
| $3d^5(^2D2)4s$ | $^1D_2$ | 271824 | 268274 | 0.9996 | 1 | -6622 |
| $3d^5(^2G1)4s$ | $^3G_5$ | 278280 | 274695 | 1.2000 | 1.2 | -6976 |
| $3d^5(^2G1)4s$ | $^3G_4$ | 278290 | 274739 | 1.0500 | 1.05 | -6928 |
| $3d^5(^2G1)4s$ | $^3G_3$ | 278290 | 274774 | 0.7507 | 0.75 | -6890 |
| $3d^5(^2G1)4s$ | $^1G_4$ | 282992 | 279200 | 1.0002 | 1 | -6774 |
| $3d^5(^4G)4p$ | $^5G_2^o$ | 284138 | 284216 | 0.3529 | 0.333 | -4971 |
| $3d^5(^4G)4p$ | $^5G_3^o$ | 284206 | 284249 | 0.8936 | 0.9166 | -4935 |
| $3d^5(^4G)4p$ | $^5G_4^o$ | 284319 | 284309 | 1.1246 | 1.15 | -4835 |
| $3d^5(^4G)4p$ | $^5G_5^o$ | 284484 | 284403 | 1.2461 | 1.266 | -4689 |
| $3d^5(^4G)4p$ | $^5G_6^o$ | 284749 | 284580 | 1.3171 | 1.333 | -4397 |
| $3d^5(^4G)4p$ | $^5H_3^o$ | 286124 | 286294 | 0.5405 | 0.5 | -5074 |
| $3d^5(^4G)4p$ | $^5H_4^o$ | 286614 | 286707 | 0.9385 | 0.9 | -4480 |
| $3d^5(^4G)4p$ | $^5H_5^o$ | 287133 | 287127 | 1.1395 | 1.1 | -4004 |
| $3d^5(^4G)4p$ | $^5H_6^o$ | 287727 | 287646 | 1.2262 | 1.214 | -3160 |
| $3d^5(^4P)4p$ | $^5D_1^o$ | 289613 | 287756 | 0.7500 | 1.5 | -4060 |
| $3d^5(^4G)4p$ | $^5F_5^o$ | 288106 | 287907 | 1.3754 | 1.4 | -4216 |
| $3d^5(^4G)4p$ | $^5F_3^o$ | 288522 | 287960 | 1.2793 | 1.25 | -4599 |
| $3d^5(^4G)4p$ | $^5H_7^o$ | 288175 | 288022 | 1.2853 | 1.285 | -2911 |
| $3d^5(^4G)4p$ | $^5F_4^o$ | 288413 | 288162 | 1.3388 | 1.35 | -4131 |
| $3d^5(^4P)4p$ | $^5S_2^o$ | 290418 | 288878 | 1.8675 | 2 | -5098 |
| $3d^5(^4G)4p$ | $^5F_1^o$ | 288825 | 289163 | 0.7410 | 0 | -5028 |
| $3d^5(^4G)4p$ | $^5F_2^o$ | 288627 | 289247 | 1.1612 | 1 | -4974 |
| $3d^5(^4P)4p$ | $^5D_0^o$ | 289135 | 290262 | | | -5601 |
| $3d^5(^4P)4p$ | $^5P_3^o$ | 292068 | 290757 | 1.5931 | 1.6667 | -4083 |
| $3d^5(^4G)4p$ | $^3F_3^o$ | 291701 | 291329 | 1.1180 | 1.0833 | -3713 |
| $3d^5(^4P)4p$ | $^5P_2^o$ | 292791 | 291390 | 1.7914 | 1.8333 | -3576 |
| $3d^5(^4P)4p$ | $^5P_1^o$ | 293024 | 291542 | 2.3610 | 2.5 | -3852 |
| $3d^5(^4G)4p$ | $^3F_4^o$ | 292101 | 291555 | 1.2688 | 1.25 | -3273 |
| $3d^5(^4G)4p$ | $^3H_6^o$ | 292235 | 291891 | 1.1699 | 1.1667 | -3681 |
| $3d^5(^4G)4p$ | $^3H_5^o$ | 292626 | 292353 | 1.0366 | 1.0333 | -3403 |
| $3d^5(^4G)4p$ | $^3H_4^o$ | 292818 | 292631 | 0.8046 | 0.8 | -3367 |
| $3d^5(^4D)4p$ | $^5F_2^o$ | 294870 | 294086 | 1.0694 | 1 | -4169 |
| $3d^5(^4D)4p$ | $^5F_3^o$ | 295332 | 294443 | 1.2708 | 1.25 | -3806 |
| $3d^5(^4D)4p$ | $^5F_4^o$ | 295953 | 294940 | 1.3604 | 1.35 | -3147 |
| $3d^5(^4D)4p$ | $^5F_5^o$ | 296412 | 295444 | 1.3983 | 1.4 | -2890 |
| $3d^5(^4D)4p$ | $^5D_3^o$ | 297672 | 296574 | 1.4912 | 1.5 | -4770 |
| $3d^5(^4G)4p$ | $^3G_3^o$ | 297427 | 296847 | 0.7556 | 0.75 | -3760 |
| $3d^5(^4G)4p$ | $^3G_4^o$ | 297564 | 296897 | 1.0547 | 1.05 | -3532 |
| $3d^5(^4D)4p$ | $^5D_4^o$ | 297899 | 296919 | 1.4841 | 1.5 | -4379 |
| $3d^5(^4G)4p$ | $^3G_5^o$ | 297689 | 296933 | 1.2018 | 1.2 | -3399 |
| $3d^5(^4D)4p$ | $^5D_2^o$ | 298147 | 297014 | 1.5342 | 1.5 | -4084 |
| $3d^5(^4D)4p$ | $^5D_1^o$ | 299204 | 297418 | 1.7653 | 1.5 | -3511 |
| $3d^5(^4D)4p$ | $^5P_1^o$ | 298499 | 297983 | 1.9948 | 2.5 | -3858 |
| $3d^5(^4P)4p$ | $^3D_1^o$ | 300187 | 298601 | 0.7187 | 0.5 | -3362 |
| $3d^5(^4D)4p$ | $^3D_3^o$ | 301604 | 298972 | 1.4074 | 1.333 | -2865 |
| $3d^5(^4D)4p$ | $^3D_2^o$ | 301654 | 300225 | 1.1698 | 1.1667 | -2944 |
| $3d^5(^4D)4p$ | $^3D_1^o$ | 302123 | 300563 | 0.5256 | 0.5 | -2333 |
| $3d^5(^4D)4p$ | $^3F_4^o$ | 302325 | 300918 | 1.2623 | 1.25 | -3443 |
| $3d^5(^4D)4p$ | $^3F_3^o$ | 303003 | 301470 | 1.1439 | 1.0833 | -2496 |

TABLE V – *Continued from previous page*

| Level | | Energy | | g-factor | | $q$ (cm$^{-1}$) |
|---|---|---|---|---|---|---|
| | | CI +$\Sigma$ | Exp. [25] | CI +$\Sigma$ | $LS$ | |
| $3d^5(^4D)4p$ | $^3F_2^o$ | 303042 | 301553 | 0.7569 | 0.6667 | -2556 |
| $3d^5(^4P)4p$ | $^3S_1^o$ | 305408 | 303250 | 1.9338 | 2 | -4071 |
| $3d^5(^2I)4p$ | $^3K_6^o$ | 305928 | 305591 | 0.9153 | 0.857 | -5915 |
| $3d^5(^4D)4p$ | $^3P_1^o$ | 307575 | 305838 | 1.5480 | 1.5 | -3562 |
| $3d^5(^2I)4p$ | $^3K_7^o$ | 306455 | 305996 | 1.0651 | 1.01785 | -5335 |
| $3d^5(^2I)4p$ | $^3I_5^o$ | 306418 | 306049 | 0.8876 | 0.8333 | -5775 |
| $3d^5(^2D1)4s$ | $^3D_3$ | 313819 | 306963 | 1.3333 | 1.333 | -7015 |
| $3d^5(^2D1)4s$ | $^3D_2$ | 313867 | 307025 | 1.1680 | 1.167 | -6897 |
| $3d^5(^2D1)4s$ | $^3D_1$ | 313924 | 307105 | 0.5090 | 0.5 | -6754 |
| $3d^5(^2I)4p$ | $^3I_6^o$ | 307816 | 307400 | 0.9987 | 1.0238 | -4316 |
| $3d^5(^2I)4p$ | $^3K_8^o$ | 308713 | 308139 | 1.1250 | 1.125 | -2908 |
| $3d^5(^2I)4p$ | $^3I_7^o$ | 308828 | 308317 | 1.0918 | 1.1428 | -2802 |
| $3d^5(^2I)4p$ | $^1H_5^o$ | 309509 | 308804 | 0.9593 | 1 | -3337 |
| $3d^5(^2I)4p$ | $^3H_6^o$ | 309855 | 309264 | 1.1352 | 1.16 | -3882 |
| $3d^5(^2I)4p$ | $^1K_7^o$ | 310317 | 309744 | 1.0063 | 1 | -3595 |
| $3d^5(^2I)4p$ | $^3H_5^o$ | 310625 | 309920 | 1.0278 | 1.0333 | -3261 |
| $3d^5(^2I)4p$ | $^3H_4^o$ | 310654 | 309953 | 0.8102 | 0.8 | -3711 |
| $3d^5(^2D_3)4p$ | $^3F_4^o$ | 312537 | 310213 | 1.2088 | 1.25 | -4007 |